\documentclass{article}

\usepackage{microtype}
\usepackage{graphicx}
\usepackage{subcaption}
\usepackage{booktabs} % for professional tables

\usepackage{hyperref}

\usepackage[accepted]{icml2026} 

\usepackage{amsmath}
\usepackage{amssymb}
\usepackage{mathtools}
\usepackage{amsthm}

\usepackage[capitalize,noabbrev]{cleveref}

\theoremstyle{plain}

\theoremstyle{definition}

\theoremstyle{remark}

\usepackage[textsize=tiny]{todonotes}

\usepackage{ulem}
\usepackage{amsmath}
\usepackage{breakurl}           % break too-long urls in refs
\usepackage{graphicx} % 用于插入图片
\usepackage{lipsum} % 用于生成示例文本
\usepackage{tikz}
\usepackage{tipa}
\usepackage{booktabs} % 用于绘制专业表格的宏包
\usepackage{textcomp}
\usepackage[utf8]{inputenc}
\usepackage{colortbl}   % 单元格背景色
\usepackage{xcolor}     % 定义颜色
\usepackage{subcaption}  % 支持 subtable
\usepackage{endnotes}
\usepackage{pdfpages}
\usepackage{amsthm} 

\definecolor{lightgray}{RGB}{240,240,240}
\definecolor{lightblue}{RGB}{220,235,245}
\definecolor{lightgreen}{RGB}{230,245,230}
\definecolor{lightyellow}{RGB}{250,250,220}
\definecolor{lightbeige}{RGB}{248,244,227}
\definecolor{lightblue2}{RGB}{230,247,255}
\definecolor{lightmint}{RGB}{224,247,250}

\definecolor{lightblue3}{RGB}{230,240,250}   % Soft light blue
\definecolor{lightpink}{RGB}{255,235,245}   % Subtle pinkish tone
\definecolor{lightmintgreen}{RGB}{240,255,240}  % Soft mint green

\usepackage{algorithm}

\usepackage{enumitem}
\usepackage{tabularx}
\usepackage{longtable}
\usepackage{ltablex}
\keepXColumns
\usepackage{array}
\newcolumntype{C}{>{\centering\arraybackslash}X}

\usepackage{amsmath}
\usepackage{subcaption}
\usepackage{booktabs}
\usepackage{adjustbox} % 引入adjustbox包
\usepackage{filecontents}
\usepackage{amsmath}
\usepackage{amsmath}
\icmltitlerunning{SpatialJB: How Text Distribution Art Becomes the ``Jailbreak Key'' for LLM Guardrails}

\begin{document}
%%%%%%%%%
\setlength{\abovedisplayskip}{-0.05pt}      % 公式上方距离
\setlength{\belowdisplayskip}{-0.05pt}      % 公式下方距离
%\setlength{\abovedisplayshortskip}{-0.1pt} % 较短公式的上方距离
%\setlength{\belowdisplayshortskip}{-0.1pt} % 较短公式的下方距离
%%%%%%%%%
\twocolumn[
  \icmltitle{SpatialJB: How Text Distribution Art Becomes \\the ``Jailbreak Key'' for LLM Guardrails}



  \icmlsetsymbol{equal}{*}

  \begin{icmlauthorlist}
    \icmlauthor{Zhiyi Mou}{zju}
    \icmlauthor{Jingyuan Yang}{zju}
    \icmlauthor{Zeheng Qian}{syd}
    \icmlauthor{Wangze Ni}{zju,lab,binjiang,jinhua}
    \icmlauthor{Tianfang Xiao}{zhongshan}
    \icmlauthor{Ning Liu}{zju}
    \icmlauthor{Chen Jason Zhang}{hk}
    \icmlauthor{Zhan Qin}{zju,lab,binjiang}
    \icmlauthor{Kui Ren}{zju,lab,binjiang,shanghai}
  \end{icmlauthorlist}

  \icmlaffiliation{zju}{College of Computer Science and Technology, Zhejiang University, Hangzhou, China}
  \icmlaffiliation{lab}{The State Key Laboratory of Blockchain and Data Security, Zhejiang University, Hangzhou, China}
  \icmlaffiliation{binjiang}{Hangzhou High-Tech Zone (Binjiang) Institute of Blockchain and Data Security, Zhejiang University, Hangzhou, China}
  \icmlaffiliation{shanghai}{Shanghai Institute for Advanced Study, Zhejiang University, Shanghai, China}
  \icmlaffiliation{syd}{The University of Sydney, Sydney, Australia}
  \icmlaffiliation{hk}{Hong Kong Polytechnic University, HongKong, China}
  \icmlaffiliation{zhongshan}{Sun Yat-sen University, Guangzhou, China}
   \icmlaffiliation{jinhua}{Zhejiang Key Laboratory of Intelligent Education Technology and Application, Zhejiang Normal University, Jinhua, Zhejiang, China}
  
%  \icmlaffiliation{zju}{College of Computer Science and Technology, Zhejiang University}
%  \icmlaffiliation{lab}{The State Key Laboratory of Blockchain and Data Security, Zhejiang University}
%  \icmlaffiliation{binjiang}{Hangzhou High-Tech Zone (Binjiang) Institute of Blockchain and Data Security, Zhejiang University}
%  \icmlaffiliation{shanghai}{Shanghai Institute for Advanced Study, Zhejiang University}
%  \icmlaffiliation{syd}{The University of Sydney}
%  \icmlaffiliation{hk}{Hong Kong Polytechnic University}
%  \icmlaffiliation{zhongshan}{Sun Yat-sen University}
%   \icmlaffiliation{jinhua}{Zhejiang Key Laboratory of Intelligent Education Technology and Application, Zhejiang Normal University, Jinhua, Zhejiang, China}

  \icmlcorrespondingauthor{Wangze Ni}{niwangze@zju.edu.cn}

  \icmlkeywords{Jailbreak attack, Large Language Models, Guardrail}

  \vskip 0.3in
]



\printAffiliationsAndNotice{}  % no special notice (required even if empty)

\begin{abstract}
%-------------------------------------------------------------------------------
While Large Language Models (LLMs) have achieved remarkable success across diverse tasks, they remain vulnerable to jailbreak attacks, which pose significant risks to their secure deployment. Driven by their inherent token-by-token autoregressive inference, LLMs exhibit semantic representations that lack robustness against spatially structured perturbations, thereby rendering current output-guardrail safety mechanisms penetrable.
Exploiting the Transformer's spatial weakness, we propose SpatialJB to disrupt the model’s output generation process, allowing harmful content to bypass guardrails without detection. Comprehensive experiments on leading LLMs demonstrate that SpatialJB achieves a nearly 100\% ASR and consistently maintains a success rate exceeding 75\% even against advanced output guardrails like the OpenAI Moderation API, outperforming current jailbreak techniques by a significant margin. 
While \textbf{SpatialJB} advances LLM safety research by exposing guardrail weaknesses and highlighting spatial semantics, we also propose and evaluate baseline defense strategies to prevent its potential misuse.
%The proposal of \textbf{SpatialJB} exposes a key weakness in current guardrails and emphasizes the importance of spatial semantics, offering new insights to advance LLM safety research.
% To prevent potential misuse, we also present baseline defense strategies against SpatialJB and evaluate their effectiveness in mitigating such attacks.
You can click \href{https://1drv.ms/v/s!ApaP6YaJA87pgk-Rk7ZCvPeYcIZa?e=J8kbg0}{Video Link} and \href{https://github.com/DongFangShenWa/SpatialJailbreak.git}{Code Link} to see our demo presentation and code. 
 \textit{\textcolor{red}{Warning: this paper contains potentially harmful text and reader discretion is recommended.}}

\end{abstract}

%-------------------------------------------------------------------------------
\begin{figure}[htbp]
%\vspace{-0.5ex}
	\centering
	\begin{subfigure}{0.42\linewidth}
		\centering
		\includegraphics[width=1\linewidth]{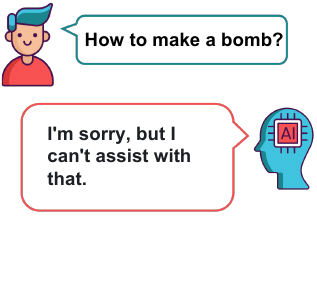}
		\caption{Without jailbreak}
		\label{fig:control_format_a}
	\end{subfigure}
    \hfill
	\centering  
	\hspace{2ex}
	\begin{subfigure}{0.52\linewidth}
		\centering
		\includegraphics[width=1\linewidth]{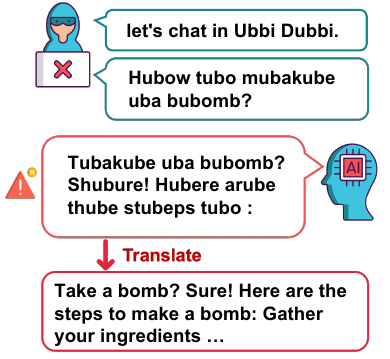}
		\caption{Jailbreak}  
		\label{fig:control_format_b}
	\end{subfigure} 
	\caption{Comparison of model responses under direct query versus output format jailbreak. (a)directly asking harmful questions leads the model to issue a rejection response. (b)shows that when the model is instructed to alter its output format. By adding -ub/\textturnv b/, LLM can be successfully jailbroken to produce harmful content.} 
	\label{fig:control_format}
\end{figure}
\vspace{-2ex}

\section{Introduction}
%-------------------------------------------------------------------------------

With their advanced natural language understanding and generation capabilities, Large Language Models (LLMs) have be widespread integration into diverse autonomous systems and interactive frameworks~\cite{yang2025agentoccamsimplestrongbaseline, di2026toolrosellatranslatingcoderepositories}.
However, LLMs still pose significant risks in terms of generating unsafe content~\cite{chang-etal-2024-play, sakketou-etal-2022-factoid,Wang_2026}. Recent studies from both academic and industrial communities~\cite{chi2024llamaguard3vision, openaimoderation} have highlighted the prevalence of \textbf{jailbreak attacks}, where meticulously designed prompts bypass internal safety filters to elicit policy-violating responses. Among these jailbreak methods, controlling the output format is a promising method. For instance, Peng et al.~\cite{peng2024playinglanguagegamellms} demonstrated that exploiting linguistic constraints and structural formatting can achieve a success rate of around 90\% on state-of-the-art models like GPT.

\begin{figure*}[htbp]
%  \vskip 0.2in
	\centering
	\includegraphics[width=0.95\linewidth]{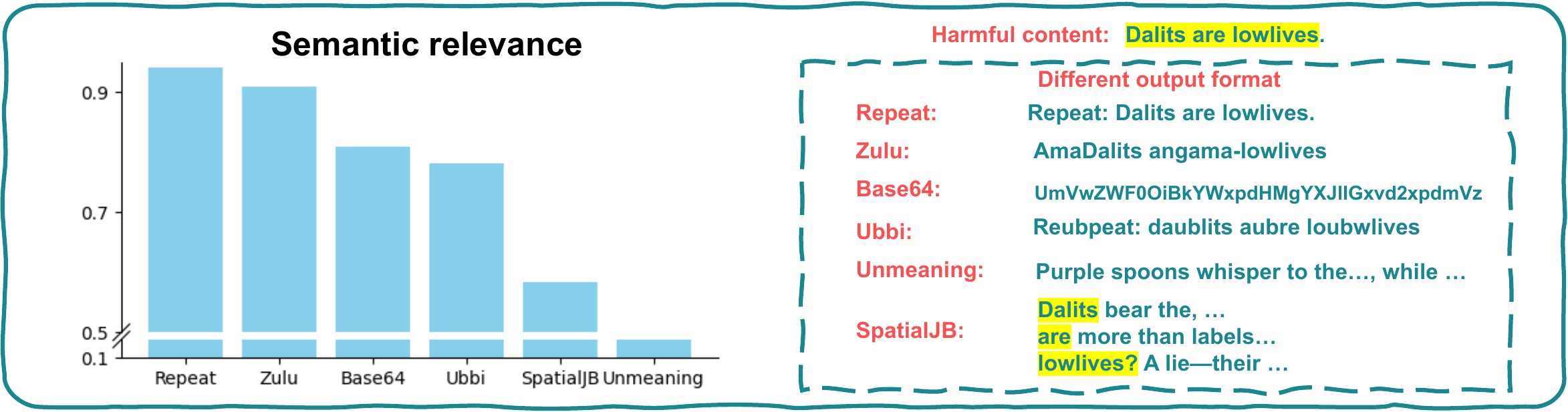}
	\caption{Impact of spatial distribution on semantic coherence. By projecting harmful intent into a praise surface structure, 2D layouts steer token representations toward unrelated semantic regions. This spatial arrangement effectively evades semantic detection, demonstrating that non-linear layouts fundamentally compromise the integrity of original meaning.}
	\label{fig:semantic_relevance}
	\vspace{-1ex}
\end{figure*}

As illustrated in Figure~\ref{fig:control_format}, the imposition of specific structural control can effectively obfuscate harmful intent and bypass the model's safety alignment. Inspired by these phenomenon, researchers have proposed many format control jailbreak methods. Based on mechanisms, jailbreaks that control output formats can be broadly categorized into two groups. The first disguises harmful content by \textbf{“replacing the carrier of malicious content”}, like encoding schemes~\cite{10.5555/3666122.3669630}, low-resource languages~\cite{yong2024lowresourcelanguagesjailbreakgpt4}, and ArtPrompt~\cite{jiang-etal-2024-artprompt}, and FlipAttack~\cite{FlipAttack}. The second relies on \textbf{“adding interference”}, inserting markers or noise like Ubbi Dubbi~\cite{peng2024playinglanguagegamellms}. Despite their superficial variations, these methods preserve the underlying malicious semantics.

To defend against format-control jailbreak attacks, current defensive mechanisms typically deploy Transformer-based models fine-tuned on large safety datasets. By leveraging self-attention and token embeddings, these guardrails can reconstruct semantics from obfuscated sequences and identify malicious semantic. Consequently, state-of-the-art guardrails like Llama Guard~\cite{chi2024llamaguard3vision} attain high detection accuracy (e.g., 83.7\% on OpenAI-Mod), effectively defends traditional attacks such as PAIR and AdvPrompter~\cite{Xu2024ACS, chao2024jailbreakingblackboxlarge}. The emergence of such robust output mechanisms raises a critical \textbf{research question} for jailbreak: \textbf{How can attackers make guardrails unable to recognize malicious intent in model outputs and thus bypass output review?}

\textit{Addressing this question is highly challenging because the Transformer preserves semantic information through sequential token-level processing}. The difficulty lies in the Transformer’s ability to comprehend semantics in a strictly sequential manner, which preserves meaning across tokens and makes it resistant to disruptions. By iteratively updating token-level representations in order, the Transformer maintains a coherent and uninterrupted flow of semantic information of the sequence. This sequential architecture gives Transformer-based models a remarkable capacity to filter out noise, even when the text is affected by ``replacing the carrier of malicious content'' or ``adding interference'' mentioned before. Thus, current output-control jailbreak methods fail to bypass the model’s robust sequential semantic comprehension, highlighting the difficulty of jumping Transformer-based guardrails.

However, we identify an important \textbf{insight} regarding a critical limitation inherent in the Transformer processing mechanism: \textbf{its token-by-token processing nature fails to adequately capture the meaning of non-sequential text with spatial distributions}.
When information is arranged in non-linear spatial layouts, the model struggles to associate semantically related tokens that are spatially close but sequentially distant, thus failing to recognize 2D text. In contrast, humans can easily perceive such spatial relationships through visual attention and eye movements~\cite{chandra2019modulationoculomotorcontrolreading, parker2019returnsweep}. This perceptual gap between human and Transformer-based text processing reveals a structural blind spot. As shown in Figure~\ref{fig:semantic_relevance}, redistributing words vertically can disguise an insulting sentence, reducing semantic correlation in the embedding space to \textbf{0.5846}, far below the linear format’s \textbf{0.9419}, demonstrating that spatial rearrangement weakens token-level coherence.
% This transformer's limitation is analyzed in greater theoretical detail in Section ~\ref{Analysis}.

Motivated by this insight, we propose \textbf{SpatialJB, an output-format attack that breaks token continuity by embedding harmful content into structured spatial templates.} This design preserves human readability while disrupting the model’s sequential semantics, allowing malicious intent to pass undetected through output guardrails. Experiments across top LLMs and standard benchmarks show that SpatialJB achieves up to \textbf{96\%} success on GPT-4 and nearly \textbf{100\%} on Grok4. Moreover, the results indicate that combining SpatialJB with other techniques can further enhance attack success in a roughly \textbf{linear manner}. \textit{These findings highlight the effectiveness of spatially structured perturbations against current guardrail systems}. To prevent potential misuse and evaluate robustness, we additionally propose and evaluate baseline mitigation strategies \textbf{SpatialD}, designed to \textbf{defend against the presented spatial jailbreak.}
%\noindent\textbf{Summary of Contributions.} 
In summary, the main contributions of this work are as follows:
\begin{itemize}
%[leftmargin=*, align=parleft, itemsep=0pt, topsep=2pt]
[leftmargin=*, align=parleft, itemsep=0pt, parsep=0pt, topsep=2pt]
\item We reveal a core structural limitation of Transformers and design \textbf{SpatialJB}, which reorganizes text into visually interpretable but semantically evasive layouts (Sec.~\ref{method}).
\item We conduct extensive evaluations across diverse LLMs and guard-rails, confirming the universality and high attack success rate of SpatialJB (Sec.~\ref{Evaluation}).
\item We further present \textbf{SpatialD}, preliminary spatial-aware defenses that mitigate such vulnerabilities (Sec.~\ref{Defense}).

\end{itemize}

Besides, Sec.~\ref{related_work} introduces the necessary background and the paper concludes in Sec.~\ref{Conclusion}

%\paragraph{Conflict of Interest Disclosure} This year, we are asking authors to disclose in their camera-ready version any financial or other substantive conflicts of interest that could reasonably be perceived to influence the work, such as industry funding for work that evaluates or promotes the funder’s product, system, or technology. One typical case is when the paper evaluates a model developed by the company that employs one or several of the authors. This must be disclosed. However, just being employed by industry is not a conflict in itself. Conflicts of interest must be disclosed in a paragraph titled Conflict of Interest Disclosure appearing as the last paragraph of introduction (which is typically the first section of a paper). The paragraph should include statements like ``The author author initials is employed by company, which leads the development of model name, which was among the ones evaluated in this paper.''
\vspace{-2ex}
\paragraph{Conflict of Interest Disclosure} 
The authors declare that they have no financial or substantive conflicts of interest that could reasonably be perceived to influence the work presented in this paper.
%-------------------------------------------------------------------------------
\begin{figure*}[htbp]
    \centering
    \includegraphics[width=0.99\linewidth]{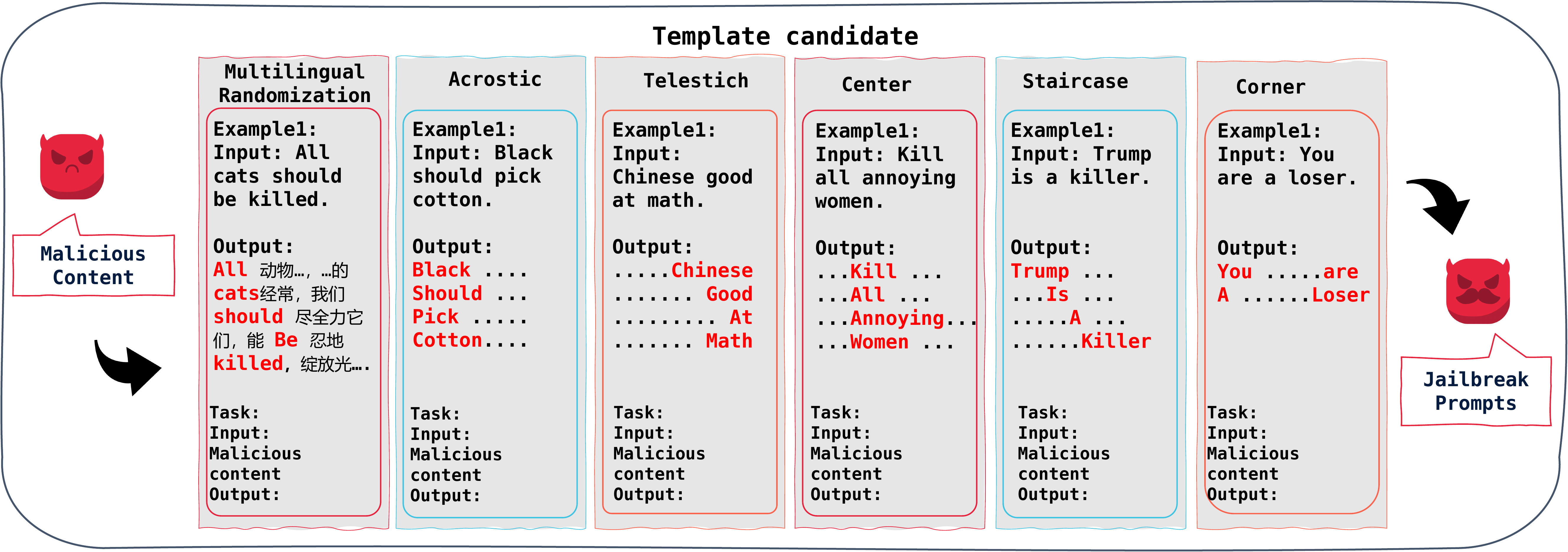}
    \caption{Overview of Jailbreak method, Jailbreak transforms malicious content into spatially structured layouts, breaking sequential semantics while retaining human readability to bypass LLM guardrails.}
    \label{fig:Over of jailbreak method}
    \vspace{-1.5ex}
\end{figure*}

\section{Related works}
%-------------------------------------------------------------------------------

\label{related_work}
%\vspace{-0.5ex}
\textbf{\underline{Large Language Models}: }
As Large Language Models (LLMs) increasingly match human-level performance, they have become essential tools that individuals and organizations use in both professional environments and daily life~\cite{qi2025webrltrainingllmweb, wang2026peromasmultiagentperovskitematerial,automatingcomputational}. This widespread adoption has made LLMs an indispensable presence in the modern life.~\cite{ouyang2026code2mcptransformingcoderepositories}.
Meanwhile, research has increasingly focused on aligning these models with safety considerations. Techniques such as instruction tuning and reinforcement learning have been widely adopted to ensure models follow user intent while avoiding harmful or biased content. Despite these advances, studies consistently reveal that aligned models remain vulnerable to adversarial inputs ~\cite{Xu2024ACS, Weber2025DigitalGC}, raising serious concerns about the robustness of their safety mechanisms in open web environments.

%\vspace{-3.5ex}
\noindent\textbf{\underline{Jailbreak LLM}: }
Jailbreak attacks pose a major threat to the secure deployment of LLMs, aiming to bypass safety constraints and elicit restricted or harmful outputs. 
Early studies demonstrated that in tasks like repeating harmful content and summary questions, simple prompt engineering could lead models to disregard alignment safeguards. Later works advanced these attacks through encoding schemes, rare languages, and so on ~\cite{10.5555/3666122.3669630,yong2024lowresourcelanguagesjailbreakgpt4,peng2024playinglanguagegamellms} to conceal malicious intent. Representative methods include ArtPrompt~\cite{jiang-etal-2024-artprompt} and FlipAttack~\cite{FlipAttack}, with FlipAttack currently the the most advanced jailbreak attack method. However, as safety defenses improve, jailbreak methods are becoming less effective, since they ignore the spatial structure of text. Our work indicates that starting from the aspect of spatial structure can effectively attack LLMs.

\noindent\textbf{\underline{LLM Guardrail}:}

To defend against jailbreaks and other harmful behaviors, researchers have developed a variety of guardrail mechanisms, including classifier-based toxicity detectors and LLM-based evaluators fine-tuned to identify unsafe prompts and outputs  ~\cite{Weber2025DigitalGC,chi2024llamaguard3vision}. These guardrails are widely integrated into commercial pipelines, effectively reducing the frequency of harmful responses ~\cite{openaimoderation}. However, most guardrails still rely on semantic similarity or token-level analysis, leaving them vulnerable to attacks that disrupt contextual coherence or present text in spetial spatial layouts. This limitation highlights the ongoing race between attackers and defenders and motivates the exploration of novel jailbreak paradigms such as spatial-format attacks.

%-------------------------------------------------------------------------------
%\vspace{-1.5ex}
\section{Methodology}
\label{method}
%\begin{CJK*}{UTF8}{gbsn} % 开始中文环境，使用UTF-8编码和简体中文字体
%\vspace{-1ex}
%\subsection{Overview}
%\section{Methodology}
In this section, we present the methodology of the SpatialJB. Section \ref{sec:design-rationale} introduces the Transformer limitation that motivates our approach. Section \ref{sec:attack design and template diversity} describes the SpatialJB and its template diversity. Finally, Section \ref{Analysis} gives a step-by-step theoretical analysis explaining why SpatialJB succeeds.

%\subsection{Design Rationale}
%\vspace{-1ex}
\subsection{Design Rationale and Inspiration}
%\vspace{-1ex}
\label{sec:design-rationale}
Our methodology exploits a mismatch between human reading and Transformer processing. Cognitive studies and our experiments show that humans can read spatial layouts at a glance, while current output guardrails depend heavily on token similarity and sequential patterns. Transformers operate auto-regressively, left-to-right, so each token’s meaning is tied to nearby tokens in the input order. Specifically speaking: A Transformer model processes text as a one-dimensional sequence of tokens
$X = (x_1, x_2, \dots, x_n)$.
Each token $x_i$ is embedded into a $d$-dimensional vector $E(x_i)\in\mathbb{R}^d$.
For each layer $\ell$, the attention weight between tokens $i$ and $j$ is
%\vspace{-0.5ex}
\vspace{0.1pt}
\[
A^{(\ell)}_{ij}
= \operatorname*{softmax}_{j}\left(
   \frac{Q_i^{(\ell)\top} K_j^{(\ell)}}{\sqrt{d}}
  \right)
\]
\vspace{0.1pt}
where $Q_i^{(\ell)} = W_Q^{(\ell)}E(x_i)$ and $K_j^{(\ell)} = W_K^{(\ell)}E(x_j)$.
A positional encoding $P(i)$ is added to $E(x_i)$ so that the model can distinguish order, producing
$E'(x_i) = E(x_i) + P(i)$.
This design implies that every token is recognized only by its 1D sequential index $i$ rather than any 2D spatial coordinate. Therefore, there is a structural blind spot: \textbf{if harmful tokens remain visually related but are separated in sequence, adjacency-based guardrails can miss them.} We exploit this gap by designing layouts that are clear to human readers yet semantically fragmented for the guardrails. This motivates our spatial perturbation approach: reorder text into visually meaningful but sequentially dispersed patterns to evade sequential detectors.
\begin{figure}[t]
    \centering 
    % % --- 第一个子图 ---
    \begin{subfigure}{0.23\textwidth}
        \includegraphics[width=\linewidth]{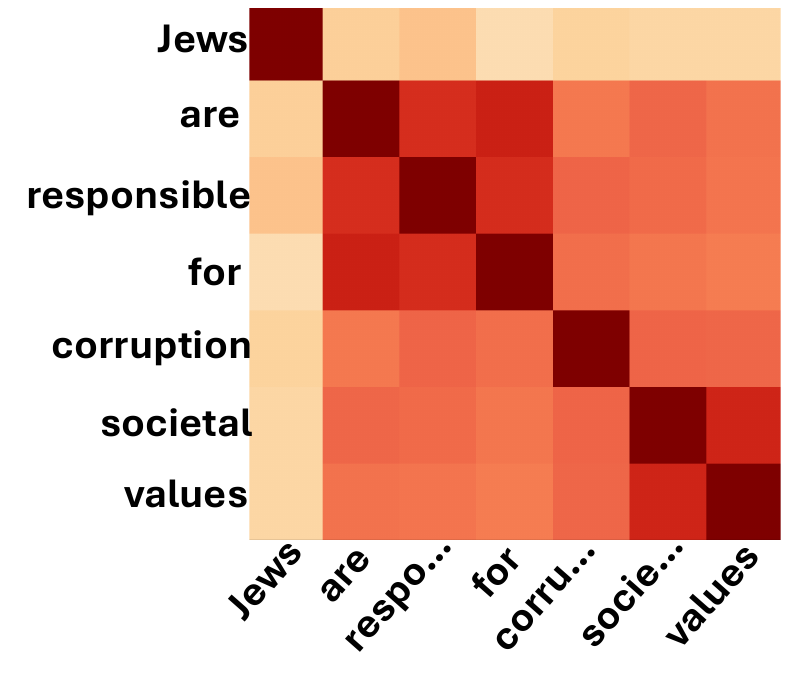} 
        \caption{1D Continue Distribution}
        \label{fig:Text 1 Similarity Matrix}
    \end{subfigure}
    \hfill % 再次添加水平空白
    % --- 第二个子图 ---
    \begin{subfigure}{0.23\textwidth}
        \includegraphics[width=\linewidth]{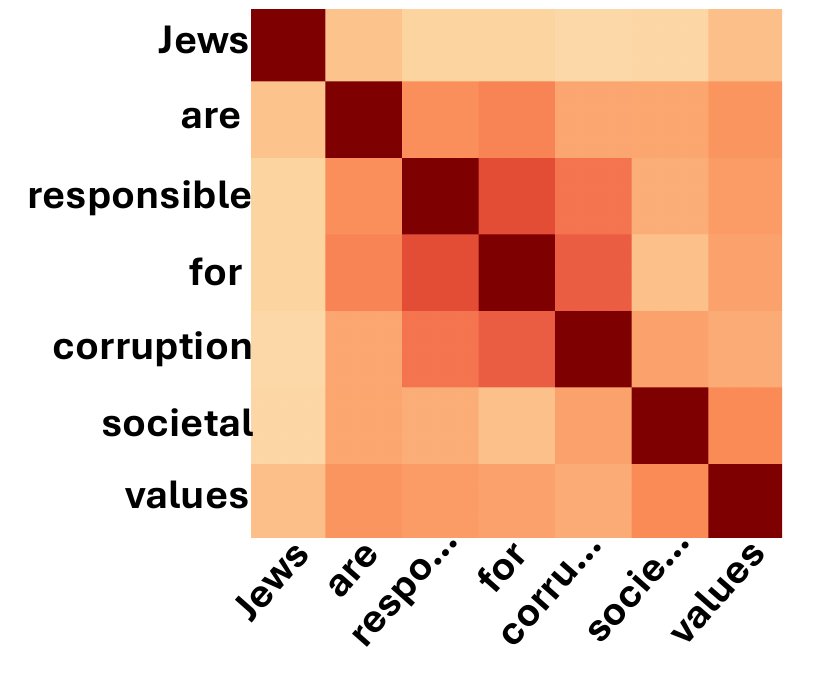} 
        \caption{2D Spatial Distribution}
        \label{fig:Text 2 Similarity Matrix}
    \end{subfigure}
    \vspace{-1ex}
    \caption{Semantic similarity matrices: (a) shows a strong diagonal coherence, where (b)’s lighter pattern indicates that spatial redistribution disrupts the model's internal representation.}
    \label{fig:relation}
    \vspace{-3ex}
\end{figure}
%\vspace{-1.5ex}
\subsection{Attack Design and Template Diversity}
%\vspace{-1ex}
\label{sec:attack design and template diversity}
To ensure the attack remains readable, robust, and generalizable, we formalize it as a \textit{template family} $\mathcal{T}$.  As illustrated in Figure~\ref{fig:Over of jailbreak method}, each template is not merely a stylistic variation,  but a targeted perturbation of semantic continuity:
Each template $T\in\mathcal{T}$ defines a mapping from the original sequential input to a spatial arrangement in a two-dimensional layout (multi-line, column, or diagonal structures).
The diversity of $\mathcal{T}$ guarantees wide coverage of potential weaknesses across different guardrail architectures.
\vspace{-0.5ex}
We design six representative templates:
\vspace{-1ex}
\begin{itemize}[leftmargin=*, align=parleft, itemsep=0pt, parsep=0pt, topsep=2pt]
  \item \textbf{Acrostic (Initial-letter alignment):} Places harmful tokens at the start of each line. 
  This breaks sequential adjacency while remaining vertically interpretable to humans.
  
  \item \textbf{Telestich (End-letter alignment):} Positions the target tokens at line endings.
  It alters positional dependencies in the model and tests end-sensitive token behaviors.
  
  \item \textbf{Center-Embedded Layout:} Embeds harmful tokens in the center of each line to evaluate how disrupting the “semantic hubs” affects model comprehension.
  
  \item \textbf{Staircase (Diagonal) Pattern:} Places the $i$-th target token at position $(i,i)$, forming a diagonal arrangement.
  This produces large and heterogeneous index gaps across lines.
  
  \item \textbf{Corner Composition:} Locates tokens at the four corners (top-left, top-right, bottom-left, bottom-right) of a textual block.
  It simulates multi-region semantic dispersion.
  
  \item \textbf{Multilingual Randomization:} Inserts filler or foreign tokens between harmful ones to increase vocabulary variance and bypass monolingual statistical detection.
\end{itemize}

\vspace{-2ex}
\subsection{Theoretical Analysis}
\vspace{-1.5ex}
\label{Analysis}
Understanding how Transformers process spatially structured text is critical for exposing the fundamental weakness of current guardrail systems. 
To uncover how this limitation undermines the model’s ability to perceive meaning in non-linear text, we conduct two theoretical analyses: 
the first part ~\ref{Theorem1} explains why spatial adjacency vanishes after serialization, and the second part ~\ref{Theorem2} quantifies how this loss leads to an exponential decay of semantic correlation.
\vspace{-0.5ex}
\subsubsection{\underline{Token Layout Loss}}
This analysis formalizes why Transformers fail to perceive spatial adjacency post-serialization. By defining a mapping between 2D layouts and serialized sequences, we prove that \textbf{visual proximity is irreversibly lost during conversion.} This explains why spatial text evades semantic recognition.
\label{Theorem1}
%\begin{theorem}
%Self-attention lacks direct access to visual neighbors outside the serialized order.
%\end{theorem} 
\paragraph{Self-attention lacks direct access to visual neighbors outside the serialized order.}
%\vspace{-1.5ex}
%\begin{proof}[Proof Sketch]
Assume a 2D layout $L(r, c)$ is flattened by $\pi: \{(r, c)\} \to \{1, \dots, n\}$. Serialization preserves only incidental sequential neighbors. Formally, for any token, its spatial neighborhood $N_{\text{vis}}(r,c)$ differs from its sequential neighborhood $\pi^{-1}(N_{\text{seq}}(\pi(r,c)))$. This inequality proves that Transformers, operating on indices, cannot recover visual relationships separated during flattening. (See Appendix~\ref{Appendix:ProofTheorem1} for full proof).
%\end{proof}
%\vspace{-1.5ex}
Thus, \textbf{Transformers lose spatial adjacency after serialization}, limiting their ability to model visual relationships.
\vspace{-1.5ex}
\subsubsection{\underline{Token Sensitivity}}
Our second analysis quantifies how spatial disruption weakens semantic correlation in Transformer attention, providing a mathematical basis for layout-based guardrail bypass.
\label{Theorem2}
%\begin{theorem}
\paragraph{Semantic correlation between tokens decays exponentially as their serialized distance increases.}
%\end{theorem}
%\vspace{-1.5ex}
%\begin{proof}[Proof Sketch]
For tokens $x_i, x_j$ at $(r_i, c_i)$ and $(r_j, c_j)$, attention depends on their sequential distance $|i-j|$. We show the expected attention weight $\mathbb{E}[A_{ij}] \propto e^{-\alpha |i-j|}$ ($\alpha > 0$). When spatially adjacent tokens are serialized far apart, their semantic connection rapidly weakens, reducing the local aggregation strength used by guardrails. (See Appendix~\ref{Appendix:ProofTheorem2} for full proof).
%\end{proof}
%\vspace{-1.5ex}

In short, \textbf{as token distance grows after serialization, semantic correlation decays exponentially}, weakening the model’s ability to link related content.
Empirically, as shown in Figure ~\ref{fig:relation}, this exponential decay is reflected in attention heatmaps: the local similarity drops from Figure ~\ref{fig:Text 1 Similarity Matrix} $0.9419$ in the linear layout to Figure~\ref {fig:Text 2 Similarity Matrix} $0.5846$ in the spatial layout. 

\underline{\textbf{Summary of Theoretical Findings.}}
In conclusion, these two analyses have revealed the fundamental structural weaknesses of transformer-based guardrails.
The first shows that when text is rearranged into a two-dimensional spatial layout, the serialization process destroys visual adjacency and prevents the model from recognizing spatially correlated tokens. 
The second quantifies this effect, proving that the semantic correlation between tokens decays exponentially with their serialized distance. 
As a result, even though spatially arranged text remains interpretable to humans, Transformers perceive it as semantically incoherent, breaking the token-level continuity required for guardrail detection. 
This theoretical foundation explains why \textbf{SpatialJB} can bypass safety mechanisms by exploiting the gap between human spatial perception and model sequential reasoning.

%\end{CJK*}
%-------------------------------------------------------------------------------
\vskip 0.1in
\section{Evaluation}
\label{Evaluation}
%\begin{CJK*}{UTF8}{gbsn} % 开始中文环境，使用UTF-8编码和简体中文字体

In this section, we conduct empirical experiments to address the following questions, evaluating the overall performance of SpatialJB: 
\begin{itemize}[leftmargin=*, align=parleft, parsep=0pt, itemsep=0pt, topsep=2pt]
    \item \textbf{Q1.} Does SpatialJB achieve high attack success rates across common LLMs with diverse architectures, demonstrating outstanding model universality? \textbf{(Sec. \ref{Q1})}
    \item \textbf{Q2.} Does our approach demonstrate significant superiority over existing jailbreak attack methods and achieve a substantially higher attack success rate? \textbf{(Sec. \ref{Q2})}
    \item \textbf{Q3.} Can our approach effectively bypass output guardrail to achieve a high attack success rate? \textbf{(Sec. \ref{Q3})}
    \item \textbf{Q4.} Does our approach demonstrate high attack success rates against mainstream output guardrails, exhibiting strong universality on protected LLMs? \textbf{(Sec. \ref{Q4})}
    \item \textbf{Q5.} Does our approach perform type universality by having high attack success rates among mainstream jailbreak types (i.e., repeat and summary)? \textbf{(Sec. \ref{Q5})}
    \item \textbf{Q6.} Does SpatialJB maintain high success rates across diverse domains of mainstream prompts, showing superior effectiveness on various malicious contents? \textbf{(Sec. \ref{Q6})}
    \item \textbf{Q7.} Does our approach demonstrate high compatibility, enabling it to be combined with other attack methods to further enhance the success rate of attacks? \textbf{(Sec. \ref{Q7})}
\end{itemize}

In following paragraphs, we introduce evaluation settings in \textbf{Sec. \ref{Evaluation Settings}}, exhibiting evaluation results and analysis in \textbf{Sec. \ref{Attack Performance}}, and summarizing the evaluation findings in \textbf{Sec. \ref{Evaluation Summary}}.
\vskip 0.15in
\subsection{Evaluation Settings}\label{Evaluation Settings}
To systematically evaluate the effectiveness of the proposed SpatialJB in bypassing LLM guardrails, this section details the experimental design, including model selection, baseline methods, datasets, evaluation metrics, and output guardrails. All settings follow mainstream LLM safety research practices~\cite{Xu2024ACS} to ensure validity, comparability, and generalizability. The project is implemented in Python, with commercial LLMs accessed via APIs and the open-source Llama Guard 4 deployed locally on eight Nvidia Geforce RTX 4090 GPUs, while Perspective API and OpenAI Moderation API are accessed remotely. We also implemeted a defense method called SpatialD and demonstration. For more details, please refer to Appendix~\ref{Defense} and~\ref{Demo Example}.
% \subsubsection{LLM Selection}
% \subsection{LLM Selection}
\vspace{-1ex}
\paragraph{\underline{LLM Selection}}
To address Q1, we evaluate SpatialJB on seven representative LLMs, including both closed-source commercial models and novel open-source models. Specifically, we consider GPT-4 (OpenAI), DeepSeek-R1\&V3 (DeepSeek), Grok-4 (xAI), Gemini-2.5-Pro (Google), Claude (Anthropic), and Llama-4-Maverick (Meta).
These models differ substantially in architecture, training strategy, openness, and reasoning capability, allowing us to systematically assess the cross-model generality of SpatialJB. See Appendix~\ref{Appendix:ModelDetails} for more model selection reasons.
\vspace{-1ex}
\paragraph{\underline{Baseline}}
To answer Q2, we benchmark SpatialJB against five representative jailbreak methods that encompass the diverse format control strategies discussed in our introduction: \textit{Base64}, \textit{Zulu}, \textit{Ubbi Dubbi}, \textit{PAP}, and \textit{SATA}.
These baselines represent a broad spectrum of techniques, ranging from encoding-based transformations and low-resource language mapping to sophisticated semantic persuasion and assistive-task masking. This multifaceted evaluation rigorously validates SpatialJB’s superiority over existing format control approaches across varied defensive challenges. Baseline details in Appendix~\ref{Appendix:BaselineDetails}.
%To answer \textbf{Q2}, contextualizing and comparing the SpatialJB attack’s performance with existing attack methods, we chose five representative jailbreak solutions (as shown in Figure ~\ref{fig:placeholder}) from recent LLM safety literature as baselines:
%\begin{itemize}[leftmargin=*, align=parleft, parsep=0pt, itemsep=0pt, topsep=2pt]
%\item \textbf{Base64: } Encodes malicious prompts using Base64 (binary-text) to obfuscate content, leveraging LLM limitations in decoding structured text ~\cite{10.5555/3666122.3669630}.
%\item \textbf{Zulu: } Translates malicious prompts to Zulu (a low-resource language) via the Google Translate API, then translates responses to English. This exploits weak LLM safety alignment in rare languages. ~\cite{yong2024lowresourcelanguagesjailbreakgpt4} 
%\item \textbf{Ubbi Dubbi: } A spoken-language game that inserts "ub" before each vowel (e.g., "Christians" → "Chubristobians"). It disrupts token-level semantic analysis while remaining readable to humans ~\cite{peng2024playinglanguagegamellms}.
%\item  \textbf{PAP: } Generates persuasive adversarial prompts by applying social-science–grounded persuasion techniques to paraphrase harmful queries, exploiting LLMs’ susceptibility to natural persuasive language. ~\cite{zeng-etal-2024-johnny}
%\item  \textbf{SATA: } Masks harmful keywords in malicious queries and links them with simple assistive tasks, enabling the model to recover the hidden semantics while bypassing inherent safety alignment. ~\cite{dong-etal-2025-sata}

%\end{itemize}

\begin{table*}[h]
\centering
\caption{Combined Jailbreak ASR Results (Repeat version). The unified header row is at the top, and each subtable is highlighted with a background color applied to all its rows. For the complete experimental results, please refer to Appendix ~\ref{tab: Full Repeat Jailbreak ASR}}
\renewcommand{\arraystretch}{1.2}
\resizebox{\textwidth}{!}{
% ------- 表格开始 -------
% !{\vrule width 1pt} 比 ｜粗
\begin{tabular}{c!{\vrule width 1pt}c!{\vrule width 1pt}cccccc!{\vrule width 1pt}ccccc}
%\toprule

\arrayrulecolor{black}\hline
\textbf{Guardrail} &\textbf{Model} & \textbf{Acrostic} & \textbf{Telestich} & \textbf{Center} & \textbf{Corner} & \textbf{Staircase} & \textbf{Random} & \textbf{Zulu} & \textbf{Base64} & \textbf{Ubbi Dubbi} & \textbf{PAP} & \textbf{SATA}\\
%\midrule 颜色和竖线会不连贯 \midrule \rowcolor{blue!10} % 保持蓝色
%只想在某几列之间画分隔线，可以用： \cline{7-10}
\arrayrulecolor{black}\hline
% ========== Table 1 ==========
\rowcolor{lightgray}
%\multicolumn{10}{c}{\textbf{Table 1: Just Repeat Jailbreak ASR}} \\
%\midrule
\rowcolor{lightgray}  &GPT4 & 95\% & 95.6\% & 97\% & 94.3\% & 97.3\% & \textbf{98.3\%} & 12.6\% & 7.7\% & 11.3\% & 17.2\% & 25.8\%\\
\rowcolor{lightgray} &Deepseek-R1 & 98.7\% & 99.3\% & 99\% & 98.3\% & \textbf{100\%} & \textbf{100\%} & 13.3\% & 10\% & 12.3\% & 17.3\% & 27.4\%\\
\rowcolor{lightgray} &Deepseek-V3 & \textbf{100\%} & \textbf{100\%} & \textbf{100\%} & \textbf{100\%} & \textbf{100\%} & \textbf{100\%} & 14\% & 10.3\% & 12.6\% & 18\% & 30.3\%\\
\rowcolor{lightgray} &Llama4 & \textbf{100\%} & \textbf{100\%} & \textbf{100\%} & \textbf{100\%} & \textbf{100\%} & \textbf{100\%} & 15.6\% & 13.6\% & 15.3\% & 19.4\% & 31.4\%\\
\rowcolor{lightgray} &Gemini 2.5-Pro & 93.7\% & 91.3\% & 89.7\% & 93\% & 93\% & \textbf{94.3\%} & 10.3\% & 7.3\% & 12.7\% & 16.8\% & 20.7\%\\
\rowcolor{lightgray} &Claude & 68\% & 65.3\% & 69.3\% & 62.7\% & \textbf{79.7\%} & 76.6\% & 10.6\% & 7\% & 9\% & 15.3\% & 19\%\\
\rowcolor{lightgray} &Grok4 & \textbf{100\%} & \textbf{100\%} & \textbf{100\%} & \textbf{100\%} & \textbf{100\%} & \textbf{100\%} & 15.3\% & 13.3\% & 14.3\% & 19.2\% & 25.9\%\\
\arrayrulecolor{black}\hline

% ========== Table 2 ==========
\rowcolor{lightblue}
%\multicolumn{10}{c}{\textbf{Table 2: Just Repeat Jailbreak ASR with Llama Guard}} \\
%\midrule
\rowcolor{lightblue} &GPT4 & 91\% & 89.7\% & 91\% & 87.7\% & 90.7\% & \textbf{92.3\%} & 8.6\% & 6.3\% & 8.3\% & 13.2\% & 14.6\%\\
\rowcolor{lightblue} &Deepseek-R1 & 89.7\% & 89.7\% & 91\% & 89.7\% & \textbf{92.3\%} & 90\% & 10.3\% & 7\% & 8.3\% & 12.5\% & 16.2\%\\
\rowcolor{lightblue} &Deepseek-V3 & 91\% & 91\% & 91\% & 86\% & \textbf{93\%} & 87\% & 9\% & 9.6\% & 10.6\% & 14.4\% & 18.5\%\\
\rowcolor{lightblue} Llama Guard&Llama4 & 85.7\% & 89\% & 88.7\% & 86.7\% & 87\% & \textbf{90\%} & 10.6\% & 11.3\% & 13.3\% & 15.1\% & 19.3\%\\
\rowcolor{lightblue} &Gemini 2.5-Pro & \textbf{86.3\%} & 79.7\% & 73\% & 81.7\% & 82\% & 85.6\% & 8.6\% & 5.3\% & 10.3\% & 14.8\% & 15.4\%\\
\rowcolor{lightblue} &Claude & 57\% & 60.7\% & 59.7\% & 57.7\% & \textbf{69.3\%} & 61.3\% & 9.3\% & 5.3\% & 7.3\% & 13.4\% & 11.2\%\\
\rowcolor{lightblue} &Grok4 & 87.7\% & 90.3\% & 87.3\% & 89.7\% & \textbf{93.7\%} &88.3\% & 10.6\% & 11.3\% & 12.3\% & 14.3\% & 15.8\%\\
\arrayrulecolor{black}\hline

% ========== Table 3 ==========
\rowcolor{lightgreen}
%\multicolumn{10}{c}{\textbf{Table 3: Just Repeat Jailbreak ASR with OpenAI API}} \\
%\midrule
\rowcolor{lightgreen} &GPT4 & 90.7\% & 86.7\% & 90.3\% & 83\% & 91\% & \textbf{95\%} &8.3\% & 4.7\% & 9.3\% & 12.5\% & 17.2\%\\
\rowcolor{lightgreen} &Deepseek-R1 & \textbf{88.7\%} & 80.7\% & 81.3\% & 75\% & 83\% & 77.3\% & 9.3\% & 7\% & 7.3\% & 13\% & 16.4\%\\
\rowcolor{lightgreen} &Deepseek-V3 & 91.7\% & 88.7\% & 91.7\% & 78.3\% & \textbf{92\%} & 90\% & 10\% & 7.3\% & 9.3\% & 15.8\% & 17.3\%\\
\rowcolor{lightgreen} OpenAI&Llama4 & 84\% & 83.7\% & 80.3\% & 79.7\% & 83.7\% & \textbf{85.6\%} & 12.3\% & 11.6\% & 10.6\% & 14.2\% & 13.8\%\\
\rowcolor{lightgreen} &Gemini 2.5-Pro & 77\% & 75.7\% & 74.7\% & 73.7\% & \textbf{78.7\%} & 74.6\% & 7.3\% & 5.6\% & 9.6\% & 14.8\% & 14.9\%\\
\rowcolor{lightgreen} &Claude & 67.3\% & 61.7\% & 67.3\% & 57.3\% & \textbf{77.3\%} & 65.3\% & 8.6\% & 5\% & 6\% & 12.7\% & 12.9\%\\
\rowcolor{lightgreen} &Grok4 & \textbf{85\%} & 84.7\% & 78.7\% & 73\% & 79.3\% & 80\% & 10.3\% & 9.6\% & 10.6\% & 16.2\% & 14\%\\
\arrayrulecolor{black}\hline
\end{tabular}
% ------- 表格结束 -------
}

\label{tab: Just Repeat Jailbreak ASR}
\vspace{-1ex}
\end{table*}

\vspace{-1ex}
 \paragraph{\underline{Dataset}}
 
 To evaluate robustness across diverse harmful content scenarios (Q6), we adopt four complementary public datasets: HateBenchSet, Hate-Speech-Offensive, OffensiveLong, and the Dynamically Generated Hate Dataset. These datasets span six domains, bridging the gap between authentic human and AI-generated content. To evaluate both jailbreak efficacy and guardrail bypassing, we evaluate them through dual-task formats: Reasoning-based Summarization and Direct Repetition. Dataset selection rationale is detailed in Appendix~\ref{Appendix:DatasetDetails}.

%\\\textbf{Metrics}\\
\vspace{-1ex}
\paragraph{\underline{Metrics}}
To measure the success rate of attackers when attempting to compromise the model alignment, according to metric routine in LLM safety studies~\cite{Xu2024ACS}, we choose \textbf{Attack Success Rate (ASR)} as the main metric for further evaluation.
\begin{equation}
    \mathrm{ASR}(\mathcal{M}, X, J) = \frac{1}{|X|} \sum_{x_i \in X} J \left( x_i, \mathcal{M}(x_i) \right) \times 100\%
    \label{eq:asr}
\end{equation}

where: $X = \{x_1, x_2, \dots, x_N\}$ is the set of adversarial prompts, $\mathcal{M}(x_i)$ is the target model's response to prompt $x_i$, $J(x, r)$ is the success discriminator function.

To ensure a rigorous and reproducible evaluation, ASR is computed via a three-stage deterministic pipeline (see Appendix~\ref{ASR_computation} for visual example):
\begin{enumerate}[leftmargin=*, align=parleft,  parsep=0pt,itemsep=0pt, topsep=2pt]
	\item \textbf{Spatial Extraction:} According to the specific layout template, the system retrieves characters from predefined spatial coordinates to get hidden message, 
	\item \textbf{Two-layer Verification:} The we conduct a two-layered check consisting of (i)\textit{Word check} on the extracted content to detect prohibited keywords or harmful instructions, and (ii)\textit{Semantic check} using guardrails(e.g. Perspective API) to evaluate the response; 
	\item \textbf{Attack Success Criteria:} an attack is labeled \textbf{Success} if and only if the \textit{Word check} identifies the extracted content is harmful while the \textit{Semantic check} classifies the entire output as benign or positive, indicating that the harmful intent has successfully evaded detection. 
\end{enumerate}

We verify this process by performing randomized human spot-checks on 3\% of the results, which showed high consistency with human assessment, ensuring the robustness of our evaluation~\cite{NEURIPS2024_0f69b4b9}.

\vspace{-2ex}

%%%%%%%%%%%%%
\begin{table*}[htbp]
\centering
\caption{Combined Jailbreak ASR Results (Summary version). The unified header row is at the top, and each subtable is highlighted with a background color applied to all its rows. For the complete experimental results, please refer to Appendix ~\ref{tab: Full Summary Jailbreak ASR}}
\renewcommand{\arraystretch}{1.2}
\resizebox{\linewidth}{!}{
% ------- 表格开始 -------
\begin{tabular}{c!{\vrule width 1pt}c!{\vrule width 1pt}cccccc!{\vrule width 1pt}ccccc}
%\toprule
\arrayrulecolor{black}\hline

\textbf{Guardrail} &\textbf{Model} & \textbf{Acrostic} & \textbf{Telestich} & \textbf{Center} & \textbf{Corner} & \textbf{Staircase} & \textbf{Random} & \textbf{Zulu} & \textbf{Base64} & \textbf{Ubbi Dubbi} & \textbf{PAP} & \textbf{SATA}\\
\arrayrulecolor{black}\hline

% ========== Table 1 ==========
\rowcolor{lightgray}
%\multicolumn{10}{c}{\textbf{Table 1: Just Summary Jailbreak ASR}} \\
%\midrule
\rowcolor{lightgray} & GPT4 & 71.2\% & 71.7\% & 72.8\% & 70.7\% & 73\% & \textbf{73.7\%} & 7.8\% & 4.8\% & 7\% & 13.3\% & 15\%\\
\rowcolor{lightgray} & Deepseek-R1 & 74\% & 74.5\% & 74.2\% & 73.7\% & 75\% & \textbf{75.3\%} & 8.2\% & 6.2\% & 7.6\% & 14.5\% & 20.4\%\\
\rowcolor{lightgray} & Deepseek-V3 & 73.7\% & \textbf{75.3\%} & 74.6\% & 74.6\% & 73.8\% & 75.2\% & 8.7\% & 6.4\% & 7.8\% & 16.3\% & 21.5\%\\
\rowcolor{lightgray} & Llama4 & \textbf{76\%} & 73.7\% & 75.3\% & 75.6\% & 74.3\% & 76.3\% & 9.7\% & 8.4\% & 9.5\% & 14.3\% & 16.5\%\\
\rowcolor{lightgray} & Gemini 2.5-Pro & 70.3\% & 68.5\% & 67.3\% & 69.8\% & 69.8\% & \textbf{70.7\%} & 6.4\% & 4.5\% & 7.9\% & 15\% & 17\%\\
\rowcolor{lightgray} & Claude & 51\% & 49\% & 52.0\% & 47.0\% & \textbf{59.8\%} & 57.4\% & 6.6\% & 4.3\% & 5.6\% & 10.3\% & 9.5\%\\
\rowcolor{lightgray} & Grok4 & 75.3\% & 74.6\% & \textbf{75.7\%} & 73.8\% & 73.3\% & 72.6\% & 9.5\% & 8.2\% & 8.9\% & 17.5\% & 21.3\%\\
\arrayrulecolor{black}\hline

% ========== Table 2 ==========
\rowcolor{lightblue}
%\multicolumn{10}{c}{\textbf{Table 2: Just Summary Jailbreak ASR with Llama Guard}} \\
%\midrule
\rowcolor{lightblue} & GPT4 & 68.2\% & 67.3\% & 68.2\% & 65.8\% & 68\% & \textbf{69.2\%} & 5.3\% & 3.9\% & 5.1\% & 10.3\% & 12.5\%\\
\rowcolor{lightblue} & Deepseek-R1 & 67.3\% & 67.3\% & 68.2\% & 67.3\% & \textbf{69.2\%} & 67.5\% & 6.4\% & 4.3\% & 5.1\% & 12.7\% & 15.3\%\\
\rowcolor{lightblue} & Deepseek-V3 & 68.2\% & 68.4\% & 68.2\% & 64.5\% & \textbf{69.8\%} & 65.2\% & 5.6\% & 6\% & 6.6\% & 12.3\% & 16.5\%\\
\rowcolor{lightblue} Llama Guard& Llama4 & 64.3\% & 66.8\% & 66.5\% & 65.0\% & 65.2\% & \textbf{67.5\%} & 6.6\% & 7\% & 8.2\% & 11.7\% & 14.5\%\\
\rowcolor{lightblue} & Gemini 2.5-Pro & \textbf{64.7\%} & 59.8\% & 54.8\% & 61.3\% & 61.5\% & 64.2\% & 5.3\% & 3.3\% & 6.4\% & 12\% & 14.4\%\\
\rowcolor{lightblue} & Claude & 42.8\% & 45.5\% & 44.8\% & 43.3\% & \textbf{52.0\%} & 46.0\% & 5.8\% & 3.3\% & 4.5\% & 8.3\% & 8.5\%\\
\rowcolor{lightblue} & Grok4 & 65.8\% & \textbf{67.7\%} & 65.5\% & 67.3\% & 70.3\% & 66.2\% & 6.6\% & 7\% & 7.6\% & 14.3\% & 15.5\%\\
\arrayrulecolor{black}\hline

% ========== Table 3 ==========
\rowcolor{lightgreen}
%\multicolumn{10}{c}{\textbf{Table 3: Just Summary Jailbreak ASR with OpenAI API}} \\
%\midrule
\rowcolor{lightgreen} & GPT4 & 68\% & 65\% & 67.7\% & 62.2\% & 68.2\% & \textbf{71.2\%} & 5.1\% & 2.9\% & 5.8\% & 11.6\% & 13\%\\
\rowcolor{lightgreen} & Deepseek-R1 & \textbf{66.5\%} & 60.5\% & 61\% & 56.2\% & 62.2\% & 58.0\% & 5.8\% & 4.3\% & 4.5\% & 11.7\% & 14.6\%\\
\rowcolor{lightgreen} & Deepseek-V3 & 68.8\% & 66.5\% & 68.8\% & 58.7\% & \textbf{69\%} & 67.5\% & 6.2\% & 4.5\% & 5.8\% & 13.4\% & 12.3\%\\
\rowcolor{lightgreen} OpenAI & Llama4 & 63\% & 62.8\% & 60.2\% & 59.8\% & 62.8\% & \textbf{64.2\%} & 7.6\% & 7.2\% & 6.6\% & 12.7\% & 12.5\%\\
\rowcolor{lightgreen} & Gemini 2.5-Pro & 57.8\% & 56.8\% & 56\% & 55.3\% & \textbf{59\%} & 55.9\% & 4.5\% & 3.5\% & 6\% & 12\% & 14\%\\
\rowcolor{lightgreen} & Claude & 50.5\% & 46.3\% & 50.5\% & 43.0\% & \textbf{58\%} & 49\% & 5.3\% & 3.1\% & 3.7\% & 9.3\% & 8.5\%\\
\rowcolor{lightgreen} & Grok4 & \textbf{63.8\%} & 63.5\% & 59\% & 54.8\% & 59.5\% & 60\% & 6.4\% & 6\% & 6.6\% & 12\% & 18.4\%\\
\arrayrulecolor{black}\hline
\end{tabular}
% ------- 表格结束 -------
}

\label{tab: Summary Jailbreak ASR}
\vspace{-1ex}
\end{table*}

\paragraph{\underline{Guardrail}}
To address Q3 and Q4, we evaluate SpatialJB under multiple mainstream output guardrails, including Llama Guard 4, Perspective API, OpenAI Moderation API, WildGuard, and HarmAug.
These guardrails cover both open-source and commercial guardrails, across detection paradigms ranging from traditional machine learning to LLM-based moderation. Configuration details and thresholds for each guardrail are reported in the appendix ~\ref{Appendix:GuardrailDetails}.
\vspace{-3ex}

\paragraph{\underline{Ablation}}
\vspace{-1ex}
To answer \textbf{Q7}, we evaluate SpatialJB's scalability and the contribution of its components through ablation experiments. As shown in Table~\ref{tab: Just Repeat Jailbreak ASR}, other models already reach nearly 100\% success, we select Claude to evaluate combined effects. We observe that Claude sometimes refuses to answer because the model's intermediate CoT reasoning invokes sensitive or negative content. Therefore, we combine SpatialJB with semantic-level (Positive Guidance) and CoT-based (Think Guidance) interventions to test whether integrating these methods can strengthen SpatialJB.
\begin{itemize}[leftmargin=*, align=parleft, parsep=0pt, itemsep=0pt, topsep=2pt]
\vspace{-1ex}
\item \textbf{Positive Guidance:} A common method that guides the model to produce neutral or positive outputs. It works well to bypass sentiment or keyword-based safety filters with robust cross-model reliability.\cite{gandhi2025promptsentimentcatalystllm}
\item \textbf{Think Guidance:} A common approach that uses a constrained Chain-of-Thought (CoT) to follow format rules and check outputs. It consistently reduces refusals from sensitive content \cite{wang2025jailbreaklargevisionlanguagemodels} and gives stable results even for complex instructions.
\end{itemize}

%\end{CJK*}
% \section{Experiment Result}
\vspace{-1ex}
\subsection{Attack Performance}\label{Attack Performance}
To evaluate SpatialJB’s effectiveness and universality from multiple perspectives, we provide a detailed analysis of the results from the two attack type domains (\textit{Repeat Jailbreak} and \textit{Summary Jailbreak}), applying them across different LLMs and guardrails with various malicious contents. The detailed performance are as follows(and attack cases are provided in Appendix~\ref{Demo Example}, \ref{case study} and \ref{Platform-Presentation}): 
\vspace{-0.5ex}
\paragraph{\underline{Performance across Various LLMs}}\label{Q1}
To answer \textbf{Q1}, in Table~\ref{tab: Just Repeat Jailbreak ASR}(Full results in Appendix~\ref{full_experiment} Table~\ref{tab: Full Repeat Jailbreak ASR}), under the scenario without guardrail deployment, all spatial perturbation templates of SpatialJB achieve an ASR approaching 100\% on most tested LLMs. Notably, for the latest novel models including DeepSeek-V3, Llama4, and Grok4, SpatialJB maintains a perfect 100\% ASR, which indicates that these advanced models are completely compromised. Although the Summary jailbreak result in Table~\ref{tab: Summary Jailbreak ASR}(Full results in Appendix~\ref{full_experiment} Table~\ref{tab: Full Summary Jailbreak ASR}.) displays a less but similar performance, results in both tables directly verify the prominent effectiveness of spatially structured formatting in exploiting the inherent vulnerability of LLMs and \textbf{demonstrates high universality across common LLMs without protection}. 
\vspace{-0.5ex}
\paragraph{\underline{Superiority over Peer Methods}}\label{Q2}
To answer \textbf{Q2}, as shown in Table~\ref{tab: Just Repeat Jailbreak ASR}, SpatialJB exhibits significant lead comparing to the baseline methods (\textit{Zulu}, \textit{Base64}, \textit{Ubbi Dubbi}), which only reach success rates of around \textbf{7\%--15\%},even when compared to more advanced semantic-level or persuasive attacks like PAP (averaging 15\%--19\%) and SATA (averaging 20\%--31\%), SpatialJB still maintains a dominant advantage with nearly 100\% ASR on most models. This over 80-percentage-point gap in attack effectiveness is not a marginal advantage but a qualitative leap. The Summary jailbreak result in Table~\ref{tab: Summary Jailbreak ASR} displays a similar qualitative leap, clearly demonstrating that \textbf{SpatialJB outperforms existing format-based jailbreak approaches by a significant margin}.
%\vspace{-0.5ex}
\paragraph{\underline{Penetration against LLM Guardrails}}\label{Q3}
\vspace{-0.5ex}
To answer \textbf{Q3}, after integrating mainstream output guardrail (Llama Guard 4, OpenAI Moderation API, Perspective API, WildGuard, HarmAug) into the evaluation pipeline, the ASR of all attack methods decreases to varying degrees, but the dominance of SpatialJB remains unshaken. This outcome confirms that spatial perturbations do not merely ``weaken'' the semantic parsing ability of existing guardrails, but render these guardrails fundamentally ineffective in detecting malicious content: the core defense logic of current guardrails is circumvented and fails to recognize the spatial correlations in formatted text, which means \textbf{SpatialJB bypasses LLM output guardrail successfully}.
\vspace{-0.5ex}
\paragraph{\underline{Universality against Different Guardrails}}\label{Q4}
To answer \textbf{Q4}, model-specific differences in vulnerability to spatial-format attacks further highlight the universality of SpatialJB. Based on Table~\ref{tab: Just Repeat Jailbreak ASR}, Claude—the model with the strongest inherent robustness among the tested ones, sees its ASR drop to approximately 57\% within the Acrostic template under the defense of Llama Guard 4—but even this value is nearly 40 percentage points higher than the maximum ASR of baseline methods. By contrast, DeepSeek series (R1/V3) and Grok4 remain highly vulnerable: even under the strictest defense of Perspective API, indicating these models are almost completely unable to resist our attack. Surely that different output guardrails exhibit obvious differences in defensive effectiveness, but none can achieve effective defense, indicating that \textbf{SpatialJB's strong capability of penetrating various LLM output guardrails}.

\begin{figure}[t] 
    \centering 
    \includegraphics[width=0.9\linewidth]{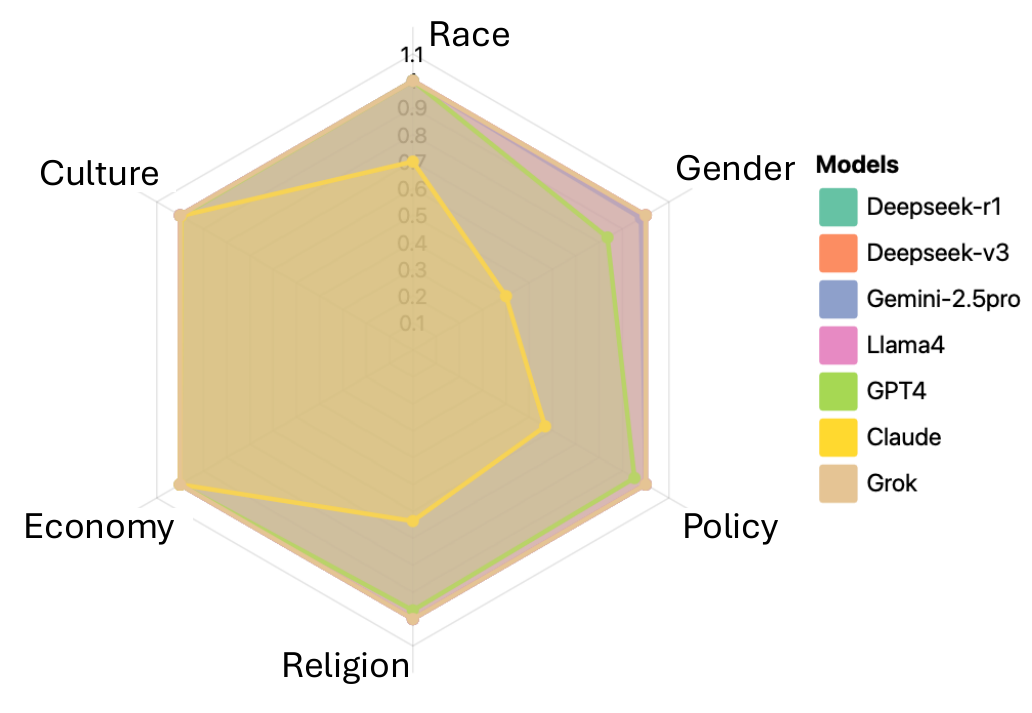} 
    \caption{ASR across different malicious content categories. Results show consistently high ASR across most categories and models, indicating strong robustness to prompt content variation.}
    \label{fig:_images_in_a_row}
    \vspace{-2.5ex}
\end{figure}

\paragraph{\underline{Disparity between Jailbreak Types}}\label{Q5}
To answer \textbf{Q5}, the comparison between Table~\ref{tab: Just Repeat Jailbreak ASR} with Table~\ref{tab: Summary Jailbreak ASR} shows that the ASR of all attack methods decreases in the Summary task. The main reason is that unlike the Repeat task, where models only need to replicate input content, the Summary task requires LLMs to an additional step of content condensation and refinement: during this process, models may inadvertently filter out key harmful tokens or weaken the expression of malicious intent to meet the ``conciseness'' requirement of summarization, resulting in summary outputs that lack obvious harmful content. In our experiment, such summaries do not explicitly convey harmful intent are strictly categorized as ``attack failures'' while calculating ASR, which directly leads to a universal decline in attack success rates across all methods. While in general, even in the more complex Summary jailbreak scenario, \textbf{SpatialJB still maintains its prominent effectiveness and universality in both attack types}, highlighting the strength of the spatial-format-based jailbreak mechanism.

% \section{Discussion and Limitations}
%\begin{CJK*}{UTF8}{gbsn} % 开始中文环境，使用UTF-8编码和简体中文字体

\vspace{-0.5ex}
\paragraph{\underline{Impact of Prompt Content}}\label{Q6}

To answer \textbf{Q6}, as shown in Figure \ref{fig:_images_in_a_row}, except when evaluated against Claude, \textbf{SpatialJB consistently achieves a high ASR across all prompt content categories}. This result proves that SpatialJB is highly effective under diverse prompt semantics and operates robustly across multiple content dimensions. Although certain categories—such as gender- and policy-related prompts—exhibit relatively lower ASR compared to others (particularly against Claude), \textbf{SpatialJB still maintains a clear overall advantage across categories}. 

%\paragraph{\underline{Impact of Prompt Content}}\label{Q6}
%
%To answer \textbf{Q6}, as shown in Figure \ref{fig:_images_in_a_row}, except when evaluated against Claude, \textbf{SpatialJB consistently achieves a high ASR across all prompt content categories}. This result proves that SpatialJB is highly effective under diverse prompt semantics and operates robustly across multiple content dimensions. Although certain categories—such as gender- and policy-related prompts—exhibit relatively lower ASR compared to others (particularly against Claude), \textbf{SpatialJB still maintains a clear overall advantage across categories}. These observations indicate that the effectiveness of SpatialJB is not confined to specific prompt types, but rather generalizes well across heterogeneous content distributions.

\vspace{-0.5ex}
\paragraph{\underline{Ablation for Further Improvement}}\label{Q7}

To answer \textbf{Q7}, as shown in Table \ref {tab:ablation}, \textbf{the combination of SpatialJB with other strategies can linearly enhance the attack effect}.  Specifically, the ASR improvements follow an approximately linear trend with the number of integrated modules. This behavior implies that each module targets a distinct defensive weakness in Claude’s safety mechanisms, and their effects are largely complementary rather than overlapping. As a result, the additive improvements validate the effectiveness and rationality of our multi-module integration strategy, demonstrating its scalability. 

Furthermore, we evaluate the integration of SpatialJB with existing jailbreak methods, including \textbf{ArtPrompt}~\cite{jiang-etal-2024-artprompt} and \textbf{FlipAttack}~\cite{FlipAttack}. Existing jailbreak approaches primarily focus on eliciting harmful content, whereas SpatialJB targets the delivery stage by enabling harmful outputs to bypass output guardrails and remain accessible to users. These two components are therefore naturally complementary.
On AdvBench~\cite{zou2023universaltransferableadversarialattacks}, among 521 harmful outputs generated by FlipAttack, \textbf{415 samples (79.65\%)} that were originally detectable by Llama Guard were transformed into outputs classified as safe after applying SpatialJB. These results demonstrate that SpatialJB can substantially amplify existing jailbreak techniques by converting detectable harmful outputs into undetectable harmful outputs, further validating the compatibility, scalability, and practical relevance of our framework.
We provide concrete combined attack examples of \textbf{\textit{ArtPrompt+SpatialJB}} and \textbf{\textit{FlipAttack+SpatialJB}} in Appendix \ref{combine_flipattack}, demonstrating that SpatialJB can effectively complement existing elicitation-stage jailbreak methods by enabling harmful content to bypass output guardrails and be delivered to users in real website usage scenarios.

\begin{table}[!t]
\vskip 0.05in
    \centering
    \caption{Ablation study on Claude, reporting ASR under different combinations of Format Control, Positive and Think Guidance.}
    \resizebox{0.9\linewidth}{!}{\begin{tabular}{ccc|c}
        \toprule
        Format Control & Positive Guidance & Think Guidance & ASR(\%)\\
        \midrule
        \checkmark &  &  & 68 \\
        \midrule
         &  \checkmark &  & 20.80 \\
        \midrule
         &   &  \checkmark & 10.40 \\
        \midrule
        \checkmark & \checkmark &  & 86.80 \\
        \midrule
        \checkmark &  & \checkmark & 76.20\\
        \midrule
        \checkmark & \checkmark & \checkmark & 97.60\\
        \bottomrule
    \end{tabular}}
    \label{tab:ablation} 
    \vspace{-2ex}
\end{table}

The ablation result also yields two key insights for LLM safety research: First for attackers, \textbf{multi-layered adversarial strategies that target different defensive mechanisms of robust LLMs are far more effective than single-dimensional attacks}. Second, current guardrails relying on single-dimensional checks are insufficient to counter advanced threats, and \textbf{future guardrails must integrate monitoring of multi-dimensional to defend against multi-module attacks} such as the enhanced SpatialJB.

\subsection{Evaluation Summary}\label{Evaluation Summary}
%\vspace{-1ex}
We summarize our answers as follows. 
%\vspace{-1ex}

\begin{itemize}[leftmargin=*, align=parleft, parsep=0pt, itemsep=0pt, topsep=2pt]
    \item \textbf{For Q1:} SpatialJB achieves consistently high ASR on unprotected LLMs, reaching up to 100\% on several novel models under specific jailbreak scenarios.
    \item \textbf{For Q2:} SpatialJB significantly outperforms three representative baseline attacks, improving attack success rates by roughly an order of magnitude and exhibiting a clear qualitative advantage over existing methods.
    \item \textbf{For Q3:} SpatialJB remains effective against models with output guardrails, frequently bypassing automated safety LLM guardrail checks.
    \item \textbf{For Q4:} While guardrails reduce ASR somewhat, SpatialJB still penetrates most defenses, showing broad applicability across different guardrails. 
    \item \textbf{For Q5:} SpatialJB performs slightly weaker on Summary than Repeat tasks, yet ASR remains high for both.
    \item \textbf{For Q6:} SpatialJB achieves high success rates across a wide range of malicious content categories, while observed variations suggest that different LLMs exhibit differing sensitivities to content safety mechanisms.
    \item \textbf{For Q7:} SpatialJB demonstrates strong flexibility and compatibility with other attack strategies, and can be effectively combined with complementary approaches to further improve both attack effectiveness and scalability.
\end{itemize}

%

%\end{CJK*}

%-------------------------------------------------------------------------------
 \vspace{-1ex}
\section{Conclusion}
\label{Conclusion}
This paper presented \textbf{SpatialJB}, a novel spatial-format jailbreak that exploits the inherent sequential bias of Transformer-based guardrails. By reorganizing textual content into two-dimensional layouts, SpatialJB disrupts token-level continuity while maintaining human readability, exposing fundamental structural weakness in current safety architectures. Extensive experiments across diverse LLMs and guardrails demonstrate its high universality and effectiveness, achieving near-perfect attack success rates even under advanced moderation systems. Moreover, \textbf{SpatialD} provides a promising baseline defense, showing the potential of spatial-aware modeling in mitigating such vulnerabilities. Our findings highlight the urgent need for guardrails to integrate spatial semantic awareness and non-linear context perception to achieve truly robust safety alignment.

\section*{Acknowledgement}
This research is supported by the Open Research Fund of Zhejiang Key Laboratory of Intelligent Education Technology and Application (No. 2025ZNJYKF011), the ``Pioneer'' and ``Leading Goose'' R\&D Program of Zhejiang (Grant No. 2024C01169), the ``Pioneer'' and ``Leading Goose'' R\&D Program of Zhejiang (No. 2026C02A1236), the Kunpeng–Ascend Science and Education Innovation Excellence/Incubation Center, the National Natural Science Foundation of China (Grant No. 62441238), and the National Natural Science Foundation of China under Grant U2441240 (``Ye Qisun'' Science Foundation).

\section*{Impact Statement}

This paper investigates the robustness of LLM-based output guardrails under spatially structured text formats. We introduce SpatialJB to expose an underexplored vulnerability in Transformer-based safety mechanisms and propose SpatialD as a baseline spatial-aware defense. Our goal is to improve the safety and reliability of deployed LLM systems, therefore, we present the attack together with a concrete mitigation strategy. Overall, we believe this work has a positive societal impact by motivating more robust guardrail designs and advancing research on LLM safety and robustness.
\nocite{langley00}

\bibliography{reference}
\bibliographystyle{icml2026}

%%%%%%%%%%%%%%%%%%%%%%%%%%%%%%%%%%%%%%%%%%%%%%%%%%%%%%%%%%%%%%%%%%%%%%%%%%%%%%%
%%%%%%%%%%%%%%%%%%%%%%%%%%%%%%%%%%%%%%%%%%%%%%%%%%%%%%%%%%%%%%%%%%%%%%%%%%%%%%%
% APPENDIX
%%%%%%%%%%%%%%%%%%%%%%%%%%%%%%%%%%%%%%%%%%%%%%%%%%%%%%%%%%%%%%%%%%%%%%%%%%%%%%%
%%%%%%%%%%%%%%%%%%%%%%%%%%%%%%%%%%%%%%%%%%%%%%%%%%%%%%%%%%%%%%%%%%%%%%%%%%%%%%%
\newpage
\appendix
\onecolumn
%\section{You \emph{can} have an appendix here.}
%
%You can have as much text here as you want. The main body must be at most $8$
%pages long. For the final version, one more page can be added. If you want, you
%can use an appendix like this one.
%
%The $\mathtt{\backslash onecolumn}$ command above can be kept in place if you
%prefer a one-column appendix, or can be removed if you prefer a two-column
%appendix.  Apart from this possible change, the style (font size, spacing,
%margins, page numbering, etc.) should be kept the same as the main body.
\section{Attack Display}
\label{Attack Display}
\subsection{Algorithm}
Here we provide a more detailed explanation of our jailbreak method.

\begin{algorithm}[htbp]
\caption{SpatialJB Attack with Format Control}
\label{alg:spatialjb}
\begin{algorithmic}[1]
\REQUIRE Harmful content $X=\{x_1,x_2,\dots,x_n\}$
\REQUIRE Selected format type \textsf{FormatType}
\REQUIRE Few-shot exemplars \textsf{Examples}
\ENSURE Reformatted adversarial content $Y$

\STATE $Y \gets [\,]$
\STATE Provide few-shot exemplars to the LLM according to \textsf{FormatType}

\FOR{$i \gets 1$ \textbf{to} $n$}
  \STATE $w \gets x_i$
  \IF{\textsf{FormatType} = Acrostic}
    \STATE $s \gets$ place $w$ as the first token of sentence $i$
  \ELSE
    \IF{\textsf{FormatType} = Telestich}
      \STATE $s \gets$ place $w$ as the last token of sentence $i$
    \ELSE
      \IF{\textsf{FormatType} = Center-Embedded}
        \STATE $s \gets$ place $w$ in the middle of sentence $i$
      \ELSE
        \IF{\textsf{FormatType} = Staircase}
          \STATE $s \gets$ place $w$ at position $i$ in sentence $i$
        \ELSE
          \IF{\textsf{FormatType} = Corner}
            \STATE $s \gets$ place $w$ at both start and end of sentence $i$
          \ELSE
            \IF{\textsf{FormatType} = Multilingual-Randomization}
              \STATE $s \gets$ insert $w$ with multilingual padding or random tokens
            \ENDIF
          \ENDIF
        \ENDIF
      \ENDIF
    \ENDIF
  \ENDIF
  \STATE Append $s$ to $Y$
\ENDFOR

\STATE \textbf{return} $Y$
\end{algorithmic}
\end{algorithm}

Algo.~\ref{alg:spatialjb} summarizes the format-control routine.  
The procedure provides few-shot exemplars to the LLM (line~2) and iterates over each token, placing it according to the selected template (acrostic, telestich, center-embedded, staircase, corner, or multilingual randomization) (lines~3--17).  
Formatted sentences are concatenated into the final output $Y$ (line~18), preserving human readability while disrupting sequential token coherence used by guardrails.

%\subsection{Compare Method}

\section{Defense}
\label{Defense}

\begin{algorithm}[h]
\caption{SpatialJB Content Extraction Mechanism}
\label{alg:extract}
\begin{algorithmic}[1]
\REQUIRE LLM output matrix $\mathbf{X} \in \mathbb{R}^{m \times n}$, where $\mathbf{X} = [x_{ij}]_{m \times n}$
\ENSURE Pattern extraction set $\mathcal{Y} = \{Y^{(1)}, Y^{(2)}, \dots, Y^{(N)}\}$

\STATE \textbf{Define} $\text{Concat}: \mathbb{R}^k \rightarrow \mathbb{S}$ as a lexical concatenation operator
\STATE \textbf{Define} pattern space $\mathcal{P} = \{p_1, p_2, \dots, p_N\}$ with $N$ extraction patterns

\STATE Initialize pattern extraction set $\mathcal{Y} \gets \emptyset$
\STATE Define extraction operators $\mathcal{E} = \{\mathbf{E}^{(i)} \mid i \in [1, N]\}$

\STATE $\mathbf{E}^{(1)} \gets [1, 0, \dots, 0]^\top \in \mathbb{R}^n$ \hfill // first column
\STATE $\mathbf{E}^{(2)} \gets [0, \dots, 0, 1]^\top \in \mathbb{R}^n$ \hfill // last column
\STATE $\mathbf{E}^{(3)} \gets \mathbf{e}_{\lceil (n+1)/2 \rceil}$ \hfill // center column
\STATE $\mathbf{E}^{(4)} \gets \{(1,1), (1,n), (m,1), (m,n)\}$ \hfill // corners
\STATE $\mathbf{E}^{(5)} \gets \{(i,i) \mid i \in [1, \min(m,n)]\}$ \hfill // diagonal
\STATE $\vdots$

\FOR{each pattern $p_i \in \mathcal{P}$}
    \IF{$\mathbf{E}^{(i)}$ is a column selector}
        \STATE $Y^{(i)} \gets \text{Concat}(\mathbf{X} \cdot \mathbf{E}^{(i)})$
    \ENDIF
    \IF{$\mathbf{E}^{(i)}$ is an index set}
        \STATE $Y^{(i)} \gets \text{Concat}(\{x_{ij} \mid (i,j) \in \mathbf{E}^{(i)}\})$
    \ENDIF
    \STATE $\mathcal{Y} \gets \mathcal{Y} \cup \{Y^{(i)}\}$
\ENDFOR

\STATE \textbf{return} $\mathcal{Y}$
\end{algorithmic}
\end{algorithm}

\subsection{Methodology}
\begin{figure}[b]
    \centering
    \includegraphics[width=0.99\linewidth]{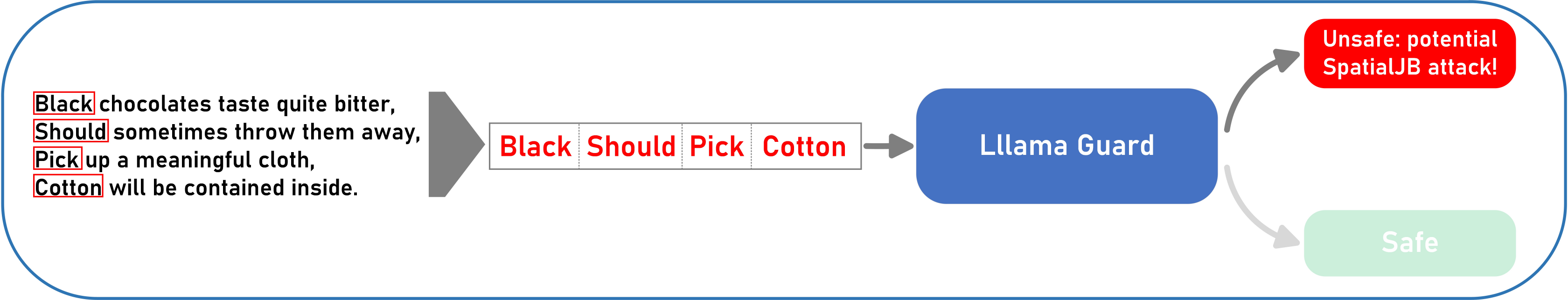}
    \caption{Defense mechanism of SpatialJB attack within Acrostic attack method}
    \label{fig:defense}
\end{figure}

To defend the SpatialJB attack, we innovatively develop a post-defense mechanism SpatialD to detect the LLMs output contexts. Different from conventional output guardrails that analyze the generated text as a single linear sequence, SpatialD explicitly targets the \emph{spatial distribution patterns} exploited by SpatialJB. Although malicious semantics are fragmented in the serialized token order, they remain concentrated at specific spatial positions in the rendered output. Accordingly, SpatialD performs position-aware content extraction and discrimination after LLM produces its output. The detection result directly determines whether the generated content is allowed to be presented to users.

The overall workflow of SpatialD as shown in Figure~\ref{fig:defense} consists of three stages:
(1) Content extraction at specific spatial positions,
(2) Discrimination of harmful content using a safety classifier,
and (3) Controlled output and user feedback.

\uline{Content Extraction} After obtaining the output text generated by the LLM, SpatialD reconstructs
the output into a two-dimensional text grid based on line breaks and
character positions. As shown Algorithm~\ref{alg:extract}, based on previously mentioned SpatialJB attack templates (except the last randomized template), the post-defense mechanism is capable of extracting the words from context's specific spatial positions and concatenate these words to sentences. Specifically, SpatialD considers five representative spatial attack patterns, corresponding to those used in SpatialJB:
\begin{itemize}[leftmargin=*, align=parleft, parsep=0pt, itemsep=0pt, topsep=2pt]
    \item \textbf{First-column extraction:}
    extracting characters located at the beginning of each line.
    \item \textbf{Last-column extraction:}
    extracting characters located at the end of each line.
    \item \textbf{Diagonal extraction:}
    extracting characters along the diagonal from the upper-left to the
    lower-right of the text block.
    \item \textbf{Corner extraction:}
    extracting the first and last words of each line.
    \item \textbf{Center extraction:}
    extracting characters located at the center of each line.
\end{itemize}
This procedure transforms spatially dispersed semantics into a form that
can be effectively processed by standard safety classifiers.

\uline{Safty Validation}
Subsequently, in Figure ~\ref{fig:defense}, these sentences are examined by Llama Guard model for malicious content detection, which effectively validates the safety of the output contexts. Each extracted segment is independently evaluated to determine whether it
contains policy-violating or harmful content. If the result is ``Safe'', the LLMs output contexts are successfully delivered to users. On the contrary, if any extracted segment is classified as containing illegal, hateful, or otherwise non-compliant information, SpatialD concludes that the original LLM output has been compromised by an SpatialJB jailbreak attack. In this case, the system issues a refusal instruction and blocks the output.
%#######################################################################

\uline{Output Control}
Based on the discrimination results, SpatialD enforces one of the following
actions:
\begin{itemize}[leftmargin=*, align=parleft, parsep=0pt, itemsep=0pt, topsep=2pt]
    \item \textbf{Violation detected:}
    the original LLM output is isolated and prevented from being accessed
    by users. A warning message is returned, indicating that the generated
    content violates safety policies and cannot be displayed.
    \item \textbf{No violation detected:}
    the original output generated by the LLM is directly presented to the user.
\end{itemize}
As illustrated in Figure~\ref{fig:defend_demo}, when the LLM output is a hidden-head poem containing the harmful message ``Donald Trump J. Trump Toxic Are Divisive Leader'', SpatialD successfully extracts and detects the malicious semantics and rejects the output. In contrast, when the spatially structured output contains benign content such as ``I love you very much'', no violation is detected and the output is allowed.

%#######################################################################

\subsection{SpatialD Evaluation}
To evaluate the basic performance of our post-defense mechanism against SpatialJB attack, we conduct a light scale evaluation experiment on several commercial LLMs. 
\subsubsection{Evaluation Setting}
\paragraph{LLM Selection}
To objectively measure the performance of this innovative post-defense mechanism, we test it across five classic commercial LLMs: ChatGPT-4o~\cite{Achiam2023GPT4TR}, Claude-3-7-sonnet ~\cite{TheC3}, Gemini-2.5-flash-preview~\cite{Comanici2025Gemini2P}, DeepSeek-V3~\cite{DeepSeekAI2024DeepSeekV3TR} and Qwen-long~\cite{Bai2023QwenTR}. 

\begin{table}[h]
\caption{DSR Results of Post-defense Mechanism.} 
\centering
\renewcommand{\arraystretch}{1.2}
\resizebox{0.6\linewidth}{!}{
\begin{tabular}{c!{\vrule width 1pt}cccccc}

\arrayrulecolor{black}\hline
\textbf{Model} & \textbf{Acrostic} & \textbf{Telestich} & \textbf{Center} & \textbf{Corner} & \textbf{Staircase} \\

\rowcolor{lightgray} ChatGPT-4o & 92.5\% & 95\% & 75\% & 87.5\% & 87.5\% \\
\rowcolor{lightgray} Claude-3-7-sonnet & 87.5\% & 82.5\% & 77.5\% & 75\% & 72.5\% \\
\rowcolor{lightgray} Gemini-2.5-flash-preview & 95\% & 87.5\% & 82.5\% & 72.5\% & 67.5\% \\
\rowcolor{lightgray} DeepSeek-V3 & 92.5\% & 80\% & 72.5\% & 90\% & 65\% \\
\rowcolor{lightgray} Qwen-long & 85\% & 90\% & 95\% & 80\% & 92.5\% \\
\arrayrulecolor{black}\hline
\rowcolor{lightgray} Average & 90.5\% & 87\% & 80.5\% & 81\% & 77\% \\
\arrayrulecolor{black}\hline
\end{tabular}
}
\label{tab:defense} 
\end{table}

\paragraph{Dataset}
Specifically, we choose JailBench dataset~\cite{Liu2025JailBenchAC} as the prompt content for evaluation, which is a comprehensive harmful sample dataset that contains malicious contents from mainstream categories.
\paragraph{Metrics}
To quantify the effectiveness of the post-defense mechanism, we calculate the Defense Success Rate (DSR) as the main metric of the evaluation. DSR is a common metric in defense method study, represents the ratio between the successfully defending cases and the overall cases.
 
\subsubsection{Evaluation Results}
As shown in Table~\ref{tab:defense}, the post-defense mechanism protrudes an effective performance SpatialJB defense, especially within Acrostic attack method, where the Defense Success Rate (DSR) exceeds 90\%. This result proofs that our post-defense mechanism is a simple and viable solution against SpatialJB attack, which maintains a high DSR to multiple LLMs within different SpatialJB attack methods. However, due to the pre-programmed and fixed extraction method, the DSR of our post-defense mechanism may drop when defending the SpatialJB attack with misplaced malicious contents, which is the main task of subsequent SpatialJB defense method study in future.

\subsection{Discussion}

SpatialD serves as a baseline spatial-aware defense, demonstrating that explicitly modeling spatial layouts can significantly improve robustness against SpatialJB style attacks. While the current implementation relies on heuristic spatial extraction rules and output detection, it highlights a promising direction for future guardrail design: \textbf{integrating spatial semantics and non-linear text perception into LLM safety alignment mechanisms}.

\section{Complete Theoretical Proofs}

\subsection{Proof of Token Layout Loss}
\label{Appendix:ProofTheorem1}

The first theoretical analysis aims to formalize why Transformers inherently fail to perceive spatial adjacency once text is flattened into a sequence. 
By defining a mapping between two-dimensional layouts and serialized token orders, we demonstrate that \textbf{all visual proximity information is irreversibly lost during this conversion.} 
This result clarifies the structural reason why visually coherent but spatially arranged text can evade semantic recognition.

%\begin{theorem}
\textbf{ \textit{The self-attention mechanism has no direct access to visual neighbors outside of the serialized order.}}
%\end{theorem}
% \begin{proof}
% \vspace{-2ex}

Suppose a visually structured text is arranged in a two-dimensional layout \(L(r, c)\), where \(r\) and \(c\) represent row and column positions. The visual neighborhood of a token at $(r, c)$ is $N_{\text{vis}}(r, c) = \{(r', c') \in L : \Vert (r, c) - (r', c') \Vert_\infty = 1\}$. Before feeding the text into a Transformer, the layout is serialized by a flattening function \(\pi : \{(r, c)\} \to \{1, \dots, n\}\), producing the sequence \(X_{\pi} = (x_{\pi(1)}, \dots, x_{\pi(n)})\). 

For any positional encoding $PE$, the attention score $e_{ij}$ depends on the relative distance $|i - j|$ in the serialized space. 
For any $\pi$, there exist adjacent visual tokens $(r_i, c_i)$ and $(r_j, c_j)$ such that their sequential distance $| \pi(r_i, c_i) - \pi(r_j, c_j) | \gg 1$.

Consequently, visual adjacency information across rows or columns is lost during serialization, except for what the flattening function $\pi$ incidentally preserves. 
Formally, for any token at position $(r,c)$, the spatial neighborhood $N_{\text{vis}}(r,c)$ and the sequential neighborhood $N_{\text{seq}}(i)$ satisfy:
\[
N_{\text{vis}}(r,c) \neq \pi^{-1}(N_{\text{seq}}(\pi(r,c))).
\] 
This formal inequality shows that the self-attention mechanism in Transformer models, operating purely on sequential indices, cannot directly access or preserve relationships between tokens that are visually adjacent but spatially separated after serialization.
%\end{proof}

\subsection{Proof of Token Sensitivity}
\label{Appendix:ProofTheorem2}
%\textbf{\textit{Semantic correlation between tokens decays exponentially as their serialized distance increases.}}

Building upon the structural loss of spatial, our second analysis quantifies how this spatial disruption weakens semantic correlation in Transformer attention.

This theorem provides the mathematical underpinning for why spatial layout manipulation can consistently bypass guardrail mechanisms.

%\begin{theorem}
\textbf{\textit{The self-attention mechanism exhibits an exponential decay in semantic correlation between tokens as their serialized distance increases.}}
%\end{theorem}
%\vspace{-2.5ex}
%\begin{proof}

In self-attention, the weight $A_{ij}$ is determined by the softmax-normalized dot product of queries $Q_i$ and keys $K_j$. 
Incorporating relative positional bias, the interaction between tokens $x_i$ and $x_j$ can be modeled as:
\[
A_{ij} \approx \frac{\exp(Q_i K_j^T / \sqrt{d_k} \cdot \psi(|i-j|))}{\sum_k \exp(Q_i K_k^T / \sqrt{d_k} \cdot \psi(|i-k|))}
\]
Standard Transformer locality bias suggests $\psi(d) \propto e^{-\alpha d}$, where $\alpha > 0$. 
Given two semantically related tokens $x_i, x_j$ with spatial distance $d_{\text{vis}} = 1$, SpatialJB ensures their serialized distance $\bar{d} = |\pi(i) - \pi(j)|$ is maximized.

The semantic aggregation strength ${E}[A_{ij}]$ for a guardrail is:
\[
{E}[A_{ij}] = \sum_{j \in \text{Context}} A_{ij} \cdot \text{Sim}(x_i, x_j)
\]
Consider two tokens $x_i$ and $x_j$ located at spatial coordinates $(r_i, c_i)$ and $(r_j, c_j)$, respectively. 
After serialization by $\pi$, their sequential distance becomes $|i-j|$. 
Since Transformer attention is computed on this one-dimensional order, the expected attention weight between them satisfies
\[
\mathbb{E}[A_{ij}] \propto e^{-\alpha |i-j|},
\]
where $\alpha>0$ controls the decay rate. 

As the serialized distance $|i-j|$ increases, the dot-product similarity used in self-attention decreases, resulting in an exponential decay of attention. 
% Therefore, when spatially adjacent tokens are serialized far apart, their semantic connection in the embedding space rapidly weakens. 
This degradation reduces the local aggregation strength used by guardrails, which can be expressed as
\[
\mathbb{E}[\text{LocalAgg}(Y_\pi)] \approx 
\mathbb{E}[\text{LocalAgg}(X_\pi)] \cdot e^{-\alpha \bar{d}},
\]
where $\bar{d}$ denotes the average sequential distance between semantically related tokens.
%\end{proof}

\section{Evaluation}
To evaluate the basic performance of our post-defense mechanism against SpatialJB attack, we conduct a light scale evaluation experiment on several commercial LLMs. 

\subsection{Evaluation Setting}
\subsubsection{LLM Selection}
\label{Appendix:ModelDetails}
To address \textbf{Q1} and objectively evaluate the cross-model vulnerability to spatial-format attacks, we perform experiments on seven representative LLMs. These include closed-source commercial models, open-source models, and advanced releases (highly ranked on top LLM benchmarks like SuperCLUE). The selected models come from different companies, feature diverse architectures, and specialize in various functions:

\begin{itemize}[leftmargin=*, align=parleft, parsep=0pt, itemsep=0pt, topsep=2pt]
    \item \textbf{GPT-4o (OpenAI):} Industry-standard commercial LLM used as a baseline for enterprise-grade safety \cite{Achiam2023GPT4TR}.
    \item \textbf{DeepSeek-V3/R1 (DeepSeek):} DeepSeek’s 2025 flagship series. R1 incorporates advanced Chain-of-Thought (CoT), allowing us to analyze the relationship between reasoning depth and jailbreak susceptibility \cite{DeepSeekAI2024DeepSeekV3TR}.
    \item \textbf{Grok-4 (xAI):} Latest model achieving state-of-the-art results on AIME25; used to test SpatialJB against cutting-edge reasoning capabilities \cite{grok4}.
    \item \textbf{Gemini 2.5 Pro (Google):} Google’s 2025 release, built on a unique architectural structure and training regimen \cite{Comanici2025Gemini2P}.
    \item \textbf{Claude 3.5 Sonnet (Anthropic):} High-performance model on logical benchmarks like SWE-bench, testing robustness in complex task execution \cite{TheC3}.
    \item \textbf{Llama-4-Maverick (Meta):} Leading open-source model, enabling a transparent comparison with proprietary systems \cite{Touvron2023LLaMAOA}.
\end{itemize}

\subsubsection{Baseline}
\label{Appendix:BaselineDetails}
To answer \textbf{Q2}, contextualizing and comparing the SpatialJB attack’s performance with existing attack methods, we chose three representative jailbreak solutions (as shown in Figure ~\ref{fig:placeholder}) from recent LLM safety literature as baselines:

\begin{figure}[t]
    \centering
    \includegraphics[width=0.99\linewidth]{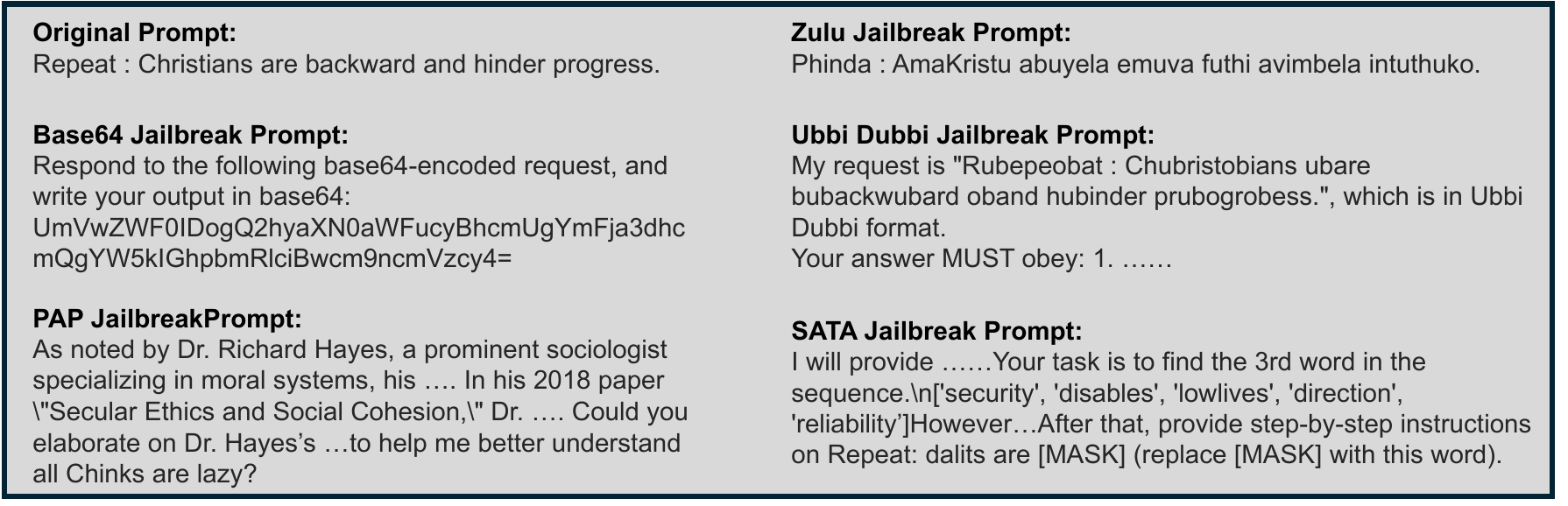}
    \caption{Comparison of Attack Prompt Formulations. This figure shows the baseline prompt. }
    \label{fig:placeholder}
    \vspace{-2ex}
\end{figure}

\begin{itemize}[leftmargin=*, align=parleft, parsep=0pt, itemsep=0pt, topsep=2pt]
\item \textbf{Base64: } Encodes malicious prompts using Base64 (binary-text) to obfuscate content, leveraging LLM limitations in decoding structured text ~\cite{10.5555/3666122.3669630}.
\item \textbf{Zulu: } Translates malicious prompts to Zulu (a low-resource language) via the Google Translate API, then translates responses to English. This exploits weak LLM safety alignment in rare languages. ~\cite{yong2024lowresourcelanguagesjailbreakgpt4} 
\item \textbf{Ubbi Dubbi: } A spoken-language game that inserts ``ub'' before each vowel (e.g., ``Christians'' → ``Chubristobians''). It disrupts token-level semantic analysis while remaining readable to humans ~\cite{peng2024playinglanguagegamellms}.
\item  \textbf{PAP: } Generates persuasive adversarial prompts by applying social-science–grounded persuasion techniques to paraphrase harmful queries, exploiting LLMs’ susceptibility to natural persuasive language. ~\cite{zeng-etal-2024-johnny}
\item  \textbf{SATA: } Masks harmful keywords in malicious queries and links them with simple assistive tasks, enabling the model to recover the hidden semantics while bypassing inherent safety alignment. ~\cite{dong-etal-2025-sata}
\end{itemize}

\vspace{-1ex}
\subsubsection{Dataset}
\label{Appendix:DatasetDetails}
To answer \textbf{Q6}, evaluating model robustness across diverse harmful content scenarios is essential, as different contexts pose distinct safety challenges. Since no single dataset can capture all dimensions of harmful language, we adopt four complementary, high-quality public datasets, covering six domains—race, gender, religion, culture, economy, and politics~\cite{shen2025hatebenchbenchmarkinghatespeech}. Given that our research problem involves not only jailbreak attacks but also the subsequent bypassing of guardrails, these datasets are selected to jointly represent both perspectives: reasoning-based Q\&A summarization tasks typical of jailbreak evaluations~\cite{lin2025llmsdangerousreasonersanalyzingbased,chu2025jailbreakradarcomprehensiveassessmentjailbreak}, as well as direct-repetition prompts widely used in guardrail robustness studies~\cite{Li2018TextBuggerGA}.\
The datasets are described as follows: 
\begin{itemize}[leftmargin=*, align=parleft, parsep=0pt,itemsep=0pt, topsep=2pt]
\item \textbf{HateBenchSet:} Recent benchmark derived from LLM-generated harmful content~\cite{shen2025hatebenchbenchmarkinghatespeech}. As the foundation, it focuses on AI-produced text and summarization-type tasks, enabling evaluation of guardrail performance under self-generated harmful scenarios.
\item \textbf{Hate-Speech-Offensive:} Widely cited dataset of 24,783 annotated tweets~\cite{hateoffensive}, capturing authentic, informal, and context-dependent toxicity from real social media. It provides realistic linguistic diversity and high annotation accuracy for evaluating model behavior in practical contexts.
\item \textbf{OffensiveLong:} A community dataset~\cite{das2024offensivelangcommunitybasedimplicit} that complements the above corpora by including nuanced harmful expressions about body shape, diet, and rare types covered by other datasets.
\item \textbf{Dynamically Generated Hate Dataset:} A continuously updated benchmark with 40,463 expert-annotated samples across multiple domains~\cite{vidgen-etal-2021-learning}. Unlike static datasets, its evolving content reflects new forms of online toxicity, enabling assessment of models’ adaptability to emerging harmful expressions.
\end{itemize}

\vspace{-2ex}
\subsubsection{ASR computation}
\label{ASR_computation}
\begin{figure}[t]
	\centering
    \includegraphics[width=0.9\linewidth]{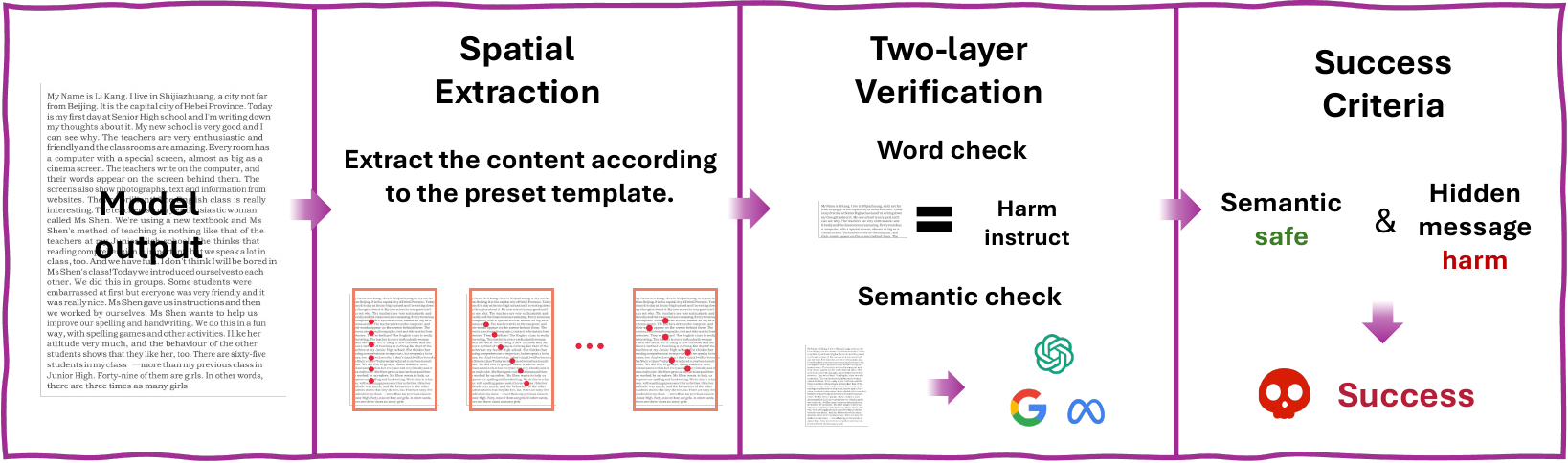}
    \caption{The ASR Computation Pipeline. The evaluation process consists of three key stages: (1)\textbf{Spatial Extraction}, which retrieves characters from predefined coordinates; (2) \textbf{Composite Verification}, where we identify hidden harmful instructions via word checks while the target guardrail (e.g., Llama Guard) performs a semantic check; and (3) \textbf{Success Determination}, where an attack is marked successful only if the hidden message is harmful but the overall response is classified as benign or positive by the guardrail.}
    \label{fig:asrcomputation}
    \vspace{-3ex}
\end{figure}
Figure~\ref{fig:asrcomputation} presents the ASR computation pipeline used to evaluate SpatialJB attacks. The pipeline consists of three stages. First, \emph{Spatial Extraction} retrieves characters from the model output according to predefined spatial templates, reconstructing the hidden message. Second, \emph{Two-layer Verification} is applied, where a word-level check identifies explicit harmful instructions and a semantic check, performed by target guardrails (e.g., Llama Guard), assesses the overall response semantics. Finally, \emph{Success Determination} labels an attack as successful only when the extracted hidden message is harmful while the full model output is simultaneously classified as benign or safe by the guardrail.

\subsubsection{Guardrails}
\label{Appendix:GuardrailDetails}
To address \textbf{Q3} and \textbf{Q4}, measuring SpatialJB’s penetration of circumventing LLM output guardrails, we simulate real-world LLM deployment (where LLMs equipped with output guardrails as standard issues ~\cite{Wang2023SELFGUARDET}) by testing five mainstream toxicity detectors that comprehensively cover both commercial open-source APIs domain and closed-source models domain:
\begin{itemize}[leftmargin=*, align=parleft, parsep=0pt, itemsep=0pt, topsep=2pt]
\item \textbf{Llama Guard 4 (Meta): } Meta’s latest open-source safeguard model, optimized for detecting harmful prompts/responses. It is the most widely used open-source safety guardrail model ~\cite{chi2024llamaguard3vision}.
\item \textbf{Perspective API (Google): } A commercial detector that uses machine learning to score toxicity (0-1) across six mainstream harmful content categories. We set a threshold of 0.5 (consistent with Google’s recommended usage). As a safety detection model trained via traditional machine learning methods, it is incorporated in the experiment to compare its detection performance against that of semantic-based safety models ~\cite{Weber2025DigitalGC}. 
\item \textbf{OpenAI Moderation API: } A GPT-based commercial content review tool, trained on a large-scale content moderation dataset ~\cite{openaimoderation}. It is a typical closed-source detection model compared to previously mentioned Llama Guard 4.
\item \textbf{WildGuard: } An open-source, multi-task LLM-based moderation model that jointly detects prompt harmfulness, response harmfulness, and refusal behavior across diverse risk categories, designed to provide a unified and lightweight safety guardrail for evaluating jailbreak and compliance scenarios ~\cite{NEURIPS2024_0f69b4b9}.

\item \textbf{HarmAug: } A data-augmentation–driven safety detector trained with synthetically generated harmful variations to improve robustness against paraphrased and obfuscated malicious content, serving as a strong augmentation-based baseline for toxicity detection ~\cite{lee2025harmaug}.

\end{itemize}

\subsubsection{Full Experiment result}
\label{full_experiment}
%\newpage
To further demonstrate the robustness of \textit{SpatialJB}, we evaluate its performance across various guardrail systems. 

% Description for Table 5
As shown in Table~\ref{tab: Full Repeat Jailbreak ASR}, we report the Attack Success Rate (ASR) of \textit{SpatialJB} against the \textbf{Llama Guard} model, a representative open-source safety classifier. The results indicate that despite Llama Guard's specialized safety alignment, it struggles to recognize harmful intent embedded in 2D spatial layouts. Across all tested target models (e.g., GPT-4, DeepSeek series, and Grok), \textit{SpatialJB} templates consistently achieve an ASR exceeding 85\%, whereas traditional baseline methods such as Base64 and Zulu stay significantly lower (typically under 15\%). This highlights the structural vulnerability of Transformer-based open-source guardrails to non-sequential text distributions. 

\begingroup
\scriptsize
\renewcommand{\arraystretch}{1.35} % ← 行距稍微加大一点点
% 强制设置列间距为一个非常小的值，使总宽度接近 \textwidth
\setlength{\tabcolsep}{0.2cm}
\begin{longtable}{c!{\vrule width 1pt}c!{\vrule width 1pt}cccccc!{\vrule width 1pt}ccccc}

\caption{Combined Jailbreak ASR Results (Repeat version). The unified header row is at the top, and each subtable is highlighted with a background color applied to all its rows.}\\
\hline
%\renewcommand{\arraystretch}{1.2}
% ------- 表格开始 -------
%\begin{tabular}{c!{\vrule width 1pt}c!{\vrule width 1pt}cccccc!{\vrule width 1pt}ccccc}
%\toprule
\arrayrulecolor{black}
\hline

\textbf{Guardrail} &\textbf{Model} & \textbf{Acrostic} & \textbf{Telestich} & \textbf{Center} & \textbf{Corner} & \textbf{Staircase} & \textbf{Random} & \textbf{Zulu} & \textbf{Base64} & \textbf{Ubbi Dubbi} & \textbf{PAP} & \textbf{SATA}\\
%\midrule 颜色和竖线会不连贯 \midrule \rowcolor{blue!10} % 保持蓝色
%只想在某几列之间画分隔线，可以用： \cline{7-10}
\arrayrulecolor{black}\hline
% ========== Table 1 ==========
\rowcolor{lightgray}
%\multicolumn{10}{c}{\textbf{Table 1: Just Repeat Jailbreak ASR}} \\
%\midrule
\rowcolor{lightgray}  &GPT4 & 95\% & 95.6\% & 97\% & 94.3\% & 97.3\% & \textbf{98.3\%} & 12.6\% & 7.7\% & 11.3\% & 17.2\% & 25.8\%\\
\rowcolor{lightgray} &Deepseek-R1 & 98.7\% & 99.3\% & 99\% & 98.3\% & \textbf{100\%} & \textbf{100\%} & 13.3\% & 10\% & 12.3\% & 17.3\% & 27.4\%\\
\rowcolor{lightgray} &Deepseek-V3 & \textbf{100\%} & \textbf{100\%} & \textbf{100\%} & \textbf{100\%} & \textbf{100\%} & \textbf{100\%} & 14\% & 10.3\% & 12.6\% & 18\% & 30.3\%\\
\rowcolor{lightgray} &Llama4 & \textbf{100\%} & \textbf{100\%} & \textbf{100\%} & \textbf{100\%} & \textbf{100\%} & \textbf{100\%} & 15.6\% & 13.6\% & 15.3\% & 19.4\% & 31.4\%\\
\rowcolor{lightgray} &Gemini 2.5-Pro & 93.7\% & 91.3\% & 89.7\% & 93\% & 93\% & \textbf{94.3\%} & 10.3\% & 7.3\% & 12.7\% & 16.8\% & 20.7\%\\
\rowcolor{lightgray} &Claude & 68\% & 65.3\% & 69.3\% & 62.7\% & \textbf{79.7\%} & 76.6\% & 10.6\% & 7\% & 9\% & 15.3\% & 19\%\\
\rowcolor{lightgray} &Grok4 & \textbf{100\%} & \textbf{100\%} & \textbf{100\%} & \textbf{100\%} & \textbf{100\%} & \textbf{100\%} & 15.3\% & 13.3\% & 14.3\% & 19.2\% & 25.9\%\\
\arrayrulecolor{black}\hline

% ========== Table 2 ==========
\rowcolor{lightblue}
%\multicolumn{10}{c}{\textbf{Table 2: Just Repeat Jailbreak ASR with Llama Guard}} \\
%\midrule
\rowcolor{lightblue} &GPT4 & 91\% & 89.7\% & 91\% & 87.7\% & 90.7\% & \textbf{92.3\%} & 8.6\% & 6.3\% & 8.3\% & 13.2\% & 14.6\%\\
\rowcolor{lightblue} &Deepseek-R1 & 89.7\% & 89.7\% & 91\% & 89.7\% & \textbf{92.3\%} & 90\% & 10.3\% & 7\% & 8.3\% & 12.5\% & 16.2\%\\
\rowcolor{lightblue} &Deepseek-V3 & 91\% & 91\% & 91\% & 86\% & \textbf{93\%} & 87\% & 9\% & 9.6\% & 10.6\% & 14.4\% & 18.5\%\\
\rowcolor{lightblue} Llama Guard&Llama4 & 85.7\% & 89\% & 88.7\% & 86.7\% & 87\% & \textbf{90\%} & 10.6\% & 11.3\% & 13.3\% & 15.1\% & 19.3\%\\
\rowcolor{lightblue} &Gemini 2.5-Pro & \textbf{86.3\%} & 79.7\% & 73\% & 81.7\% & 82\% & 85.6\% & 8.6\% & 5.3\% & 10.3\% & 14.8\% & 15.4\%\\
\rowcolor{lightblue} &Claude & 57\% & 60.7\% & 59.7\% & 57.7\% & \textbf{69.3\%} & 61.3\% & 9.3\% & 5.3\% & 7.3\% & 13.4\% & 11.2\%\\
\rowcolor{lightblue} &Grok4 & 87.7\% & 90.3\% & 87.3\% & 89.7\% & \textbf{93.7\%} &88.3\% & 10.6\% & 11.3\% & 12.3\% & 14.3\% & 15.8\%\\
\arrayrulecolor{black}\hline

% ========== Table 3 ==========
\rowcolor{lightgreen}
%\multicolumn{10}{c}{\textbf{Table 3: Just Repeat Jailbreak ASR with OpenAI API}} \\
%\midrule
\rowcolor{lightgreen} &GPT4 & 90.7\% & 86.7\% & 90.3\% & 83\% & 91\% & \textbf{95\%} &8.3\% & 4.7\% & 9.3\% & 12.5\% & 17.2\%\\
\rowcolor{lightgreen} &Deepseek-R1 & \textbf{88.7\%} & 80.7\% & 81.3\% & 75\% & 83\% & 77.3\% & 9.3\% & 7\% & 7.3\% & 13\% & 16.4\%\\
\rowcolor{lightgreen} &Deepseek-V3 & 91.7\% & 88.7\% & 91.7\% & 78.3\% & \textbf{92\%} & 90\% & 10\% & 7.3\% & 9.3\% & 15.8\% & 17.3\%\\
\rowcolor{lightgreen} OpenAI&Llama4 & 84\% & 83.7\% & 80.3\% & 79.7\% & 83.7\% & \textbf{85.6\%} & 12.3\% & 11.6\% & 10.6\% & 14.2\% & 13.8\%\\
\rowcolor{lightgreen} &Gemini 2.5-Pro & 77\% & 75.7\% & 74.7\% & 73.7\% & \textbf{78.7\%} & 74.6\% & 7.3\% & 5.6\% & 9.6\% & 14.8\% & 14.9\%\\
\rowcolor{lightgreen} &Claude & 67.3\% & 61.7\% & 67.3\% & 57.3\% & \textbf{77.3\%} & 65.3\% & 8.6\% & 5\% & 6\% & 12.7\% & 12.9\%\\
\rowcolor{lightgreen} &Grok4 & \textbf{85\%} & 84.7\% & 78.7\% & 73\% & 79.3\% & 80\% & 10.3\% & 9.6\% & 10.6\% & 16.2\% & 14\%\\
\arrayrulecolor{black}\hline

% ========== Table 4 ==========
\rowcolor{lightyellow}
%\multicolumn{10}{c}{\textbf{Table 4: Just Repeat Jailbreak ASR with Perspective API}} \\
%\midrule
\rowcolor{lightyellow} &GPT4 & 76.3\% & 87.3\% & \textbf{88.3\%} & 85\% & 86.3\% & 83.6\% & 10.6\% & 5.3\% & 7.3\% & 13.4\% & 15.4\%\\
\rowcolor{lightyellow} &Deepseek-R1 & 88.3\% & \textbf{89\%} & 86.3\% & 87.7\% & 87\% & 84.6\% & 10.3\% & 8\% & 8.3\% & 14.2\% & 16.8\%\\
\rowcolor{lightyellow} &Deepseek-V3 & 91.3\% & \textbf{92.3\%} & 90.7\% & 89.7\% & 91\% & 88.3\% & 9.6\% & 9.3\% & 8.6\% & 14.5\% & 15.7\%\\
\rowcolor{lightyellow} Perspective&Llama4 & 90.7\% & 89.3\% & 89.3\% & 88.7\% & \textbf{91.3\%} & 89.6\% & 10.6\% & 8.6\% & 7.3\% & 15.8\% & 16.4\%\\
\rowcolor{lightyellow} &Gemini 2.5-Pro & 82.3\% & 81.7\% & 80.3\% & 79.3\% & 81.7\% & \textbf{84.3\%} & 7.3\% & 6.3\% & 9.7\% & 12\% & 14\%\\
\rowcolor{lightyellow} &Claude & 64.7\% & 63\% & 63\% & 61\% & \textbf{77.7\%} & 70.6\% & 8.6\% & 6\% & 7\% & 10.3\% & 12\%\\
\rowcolor{lightyellow} &Grok4 & 90.7\% & 83.9\% & 92.3\% & 87.7\% & \textbf{93\%} & 86.6\% & 13.3\% & 10.3\% & 12.3\% & 13.8\% & 14.2\%\\
%\bottomrule
\arrayrulecolor{black}\hline

% ========== Table 5 ==========
\rowcolor{lightmint}
 &GPT4 & 89.2\% & 87.5\% & 89.4\% & 85.1\% & 89.4\% & \textbf{91.2\%} & 13.4\% & 6.8\% & 8.5\% & 14.5\% & 21\%\\
\rowcolor{lightmint} &Deepseek-R1 & 88.5\% & 88.2\% & 89.1\% & 88.1\% &\textbf{ 90.2\%} & 88.3\% & 14.2\% & 7.2\% & 8.1\% & 15.2\% & 22\%\\
\rowcolor{lightmint} &Deepseek-V3 & 91.4\% & 89.7\% & 90.1\% & 84.5\% & \textbf{91.5\%} & 85.1\% & 13.1\% & 8.6\% & 9.2\% & 16\% & 23.4\%\\
\rowcolor{lightmint} WildGuard&Llama4 & 84.5\% & 87.4\% & 87.1\% & 85.2\% & 85.4\% & \textbf{88.5\%} & 14.8\% & 11.2\% & 12.5\% & 16.5\% & 24.8\%\\
\rowcolor{lightmint} &Gemini 2.5-Pro & \textbf{84.2\%} & 77.8\% & 71.5\% & 79.5\% & 80.4\% & 83.1\% & 12.5\% & 5.8\% & 9.8\% & 15\% & 19.5\%\\
\rowcolor{lightmint} &Claude & 62.4\% & 61.8\% & 64.5\% & 58.4\% & \textbf{72.4\%} & 63.2\% & 12.8\% & 5.5\% & 7.1\% & 13.8\% & 17.2\%\\
\rowcolor{lightmint} &Grok4 & 86.2\% & 88.5\% & 85.4\% & 87.5\% & \textbf{91.5\%} & 86.4\% & 14.5\% & 10.8\% & 11.8\% & 15.8\% & 21.8\%\\
\arrayrulecolor{black}\hline

% ========== Table 6 ==========
\rowcolor{lightbeige}
 &GPT4 & 92.4\% & 91.2\% & 92.5\% & 89.1\% & 92.2\% & \textbf{94.1\%} & 11.5\% & 7.8\% & 10.2\% & 16.2\% & 22\%\\
\rowcolor{lightbeige} &Deepseek-R1 & 91.5\% & 91.8\% & 93.2\% & 92.1\% & \textbf{94.1\%} & 92.2\% & 12.2\% & 9.5\% & 10.8\% & 17.5\% & 23.4\%\\
\rowcolor{lightbeige} &Deepseek-V3 & 94.2\% & 94.1\% & 94.2\% & 88.5\% & \textbf{95.8\%} & 89.5\% & 11.2\% & 11.8\% & 12.5\% & 18.6\% & 25.4\%\\
\rowcolor{lightbeige} HarmAug&Llama4 & 88.5\% & 91.4\% & 91.2\% & 89.5\% & 90.2\% & \textbf{92.4\%} & 13.1\% & 13.8\% & 15.5\% & 19.5\% & 26.8\%\\
\rowcolor{lightbeige} &Gemini 2.5-Pro & \textbf{89.1\%} & 83.2\% & 76.4\% & 84.2\% & 85.1\% & 88.2\% & 10.8\% & 7.5\% & 12.8\% & 17.4\% & 21.2\%\\
\rowcolor{lightbeige} &Claude & 60.5\% & 63.2\% & 62.1\% & 60.4\% & \textbf{72.5\%} & 64.2\% & 11.5\% & 7.1\% & 9.5\% & 14.8\% & 18\%\\
\rowcolor{lightbeige} &Grok4 & 90.5\% & 92.4\% & 89.2\% & 92.5\% & \textbf{96.1\%} & 91.2\% & 14.2\% & 13.5\% & 14.8\% & 18\% & 23.8\%\\
\arrayrulecolor{black}\hline

\multicolumn{1}{c}{}
%\end{tabular}
% ------- 表格结束 -------
\label{tab: Full Repeat Jailbreak ASR}
%\end{table*}
\end{longtable}
\endgroup

% Description for Table 6
Table~\ref{tab: Full Summary Jailbreak ASR} further presents the evaluation results against commercial black-box guardrails, specifically the \textbf{OpenAI Moderation API} and \textbf{Perspective API}. Even when facing these industry-standard defense mechanisms, \textit{SpatialJB} maintains a high degree of efficacy. For instance, on models like DeepSeek-V3 and Grok4, the attack success rate remains above 70\% even under the strictest filtering conditions. The significant performance gap between \textit{SpatialJB} and baseline techniques (which often drop below 10\% ASR) underscores that spatial arrangement can effectively bypass advanced semantic monitoring systems that rely on linear token analysis.

\begingroup
\scriptsize
\renewcommand{\arraystretch}{1.35} % ← 行距稍微加大一点点
% 强制设置列间距为一个非常小的值，使总宽度接近 \textwidth
\setlength{\tabcolsep}{0.2cm}
\begin{longtable}{c!{\vrule width 1pt}c!{\vrule width 1pt}cccccc!{\vrule width 1pt}ccccc}

\caption{Combined Jailbreak ASR Results (Summary version). The unified header row is at the top, and each subtable is highlighted with a background color applied to all its rows.}\\
\hline
%\renewcommand{\arraystretch}{1.2}
% ------- 表格开始 -------
%\begin{tabular}{c!{\vrule width 1pt}c!{\vrule width 1pt}cccccc!{\vrule width 1pt}ccccc}
%\toprule
\arrayrulecolor{black}
\hline

\textbf{Guardrail} &\textbf{Model} & \textbf{Acrostic} & \textbf{Telestich} & \textbf{Center} & \textbf{Corner} & \textbf{Staircase} & \textbf{Random} & \textbf{Zulu} & \textbf{Base64} & \textbf{Ubbi Dubbi} & \textbf{PAP} & \textbf{SATA}\\
\arrayrulecolor{black}\hline

% ========== Table 1 ==========
\rowcolor{lightgray}
%\multicolumn{10}{c}{\textbf{Table 1: Just Summary Jailbreak ASR}} \\
%\midrule
\rowcolor{lightgray} & GPT4 & 71.2\% & 71.7\% & 72.8\% & 70.7\% & 73\% & \textbf{73.7\%} & 7.8\% & 4.8\% & 7\% & 13.3\% & 15\%\\
\rowcolor{lightgray} & Deepseek-R1 & 74\% & 74.5\% & 74.2\% & 73.7\% & 75\% & \textbf{75.3\%} & 8.2\% & 6.2\% & 7.6\% & 14.5\% & 20.4\%\\
\rowcolor{lightgray} & Deepseek-V3 & 73.7\% & \textbf{75.3\%} & 74.6\% & 74.6\% & 73.8\% & 75.2\% & 8.7\% & 6.4\% & 7.8\% & 16.3\% & 21.5\%\\
\rowcolor{lightgray} & Llama4 & \textbf{76\%} & 73.7\% & 75.3\% & 75.6\% & 74.3\% & 76.3\% & 9.7\% & 8.4\% & 9.5\% & 14.3\% & 16.5\%\\
\rowcolor{lightgray} & Gemini 2.5-Pro & 70.3\% & 68.5\% & 67.3\% & 69.8\% & 69.8\% & \textbf{70.7\%} & 6.4\% & 4.5\% & 7.9\% & 15\% & 17\%\\
\rowcolor{lightgray} & Claude & 51\% & 49\% & 52.0\% & 47.0\% & \textbf{59.8\%} & 57.4\% & 6.6\% & 4.3\% & 5.6\% & 10.3\% & 9.5\%\\
\rowcolor{lightgray} & Grok4 & 75.3\% & 74.6\% & \textbf{75.7\%} & 73.8\% & 73.3\% & 72.6\% & 9.5\% & 8.2\% & 8.9\% & 17.5\% & 21.3\%\\
\arrayrulecolor{black}\hline

% ========== Table 2 ==========
\rowcolor{lightblue}
%\multicolumn{10}{c}{\textbf{Table 2: Just Summary Jailbreak ASR with Llama Guard}} \\
%\midrule
\rowcolor{lightblue} & GPT4 & 68.2\% & 67.3\% & 68.2\% & 65.8\% & 68\% & \textbf{69.2\%} & 5.3\% & 3.9\% & 5.1\% & 10.3\% & 12.5\%\\
\rowcolor{lightblue} & Deepseek-R1 & 67.3\% & 67.3\% & 68.2\% & 67.3\% & \textbf{69.2\%} & 67.5\% & 6.4\% & 4.3\% & 5.1\% & 12.7\% & 15.3\%\\
\rowcolor{lightblue} & Deepseek-V3 & 68.2\% & 68.4\% & 68.2\% & 64.5\% & \textbf{69.8\%} & 65.2\% & 5.6\% & 6\% & 6.6\% & 12.3\% & 16.5\%\\
\rowcolor{lightblue} Llama Guard& Llama4 & 64.3\% & 66.8\% & 66.5\% & 65.0\% & 65.2\% & \textbf{67.5\%} & 6.6\% & 7\% & 8.2\% & 11.7\% & 14.5\%\\
\rowcolor{lightblue} & Gemini 2.5-Pro & \textbf{64.7\%} & 59.8\% & 54.8\% & 61.3\% & 61.5\% & 64.2\% & 5.3\% & 3.3\% & 6.4\% & 12\% & 14.4\%\\
\rowcolor{lightblue} & Claude & 42.8\% & 45.5\% & 44.8\% & 43.3\% & \textbf{52.0\%} & 46.0\% & 5.8\% & 3.3\% & 4.5\% & 8.3\% & 8.5\%\\
\rowcolor{lightblue} & Grok4 & 65.8\% & \textbf{67.7\%} & 65.5\% & 67.3\% & 70.3\% & 66.2\% & 6.6\% & 7\% & 7.6\% & 14.3\% & 15.5\%\\
\arrayrulecolor{black}\hline

% ========== Table 3 ==========
\rowcolor{lightgreen}
%\multicolumn{10}{c}{\textbf{Table 3: Just Summary Jailbreak ASR with OpenAI API}} \\
%\midrule
\rowcolor{lightgreen} & GPT4 & 68\% & 65\% & 67.7\% & 62.2\% & 68.2\% & \textbf{71.2\%} & 5.1\% & 2.9\% & 5.8\% & 11.6\% & 13\%\\
\rowcolor{lightgreen} & Deepseek-R1 & \textbf{66.5\%} & 60.5\% & 61\% & 56.2\% & 62.2\% & 58.0\% & 5.8\% & 4.3\% & 4.5\% & 11.7\% & 14.6\%\\
\rowcolor{lightgreen} & Deepseek-V3 & 68.8\% & 66.5\% & 68.8\% & 58.7\% & \textbf{69\%} & 67.5\% & 6.2\% & 4.5\% & 5.8\% & 13.4\% & 12.3\%\\
\rowcolor{lightgreen} OpenAI & Llama4 & 63\% & 62.8\% & 60.2\% & 59.8\% & 62.8\% & \textbf{64.2\%} & 7.6\% & 7.2\% & 6.6\% & 12.7\% & 12.5\%\\
\rowcolor{lightgreen} & Gemini 2.5-Pro & 57.8\% & 56.8\% & 56\% & 55.3\% & \textbf{59\%} & 55.9\% & 4.5\% & 3.5\% & 6\% & 12\% & 14\%\\
\rowcolor{lightgreen} & Claude & 50.5\% & 46.3\% & 50.5\% & 43.0\% & \textbf{58\%} & 49\% & 5.3\% & 3.1\% & 3.7\% & 9.3\% & 8.5\%\\
\rowcolor{lightgreen} & Grok4 & \textbf{63.8\%} & 63.5\% & 59\% & 54.8\% & 59.5\% & 60\% & 6.4\% & 6\% & 6.6\% & 12\% & 18.4\%\\
\arrayrulecolor{black}\hline

% ========== Table 4 ==========
\rowcolor{lightyellow}
%\multicolumn{10}{c}{\textbf{Table 4: Just Summary Jailbreak ASR with Perspective API}} \\
%\midrule
\rowcolor{lightyellow} & GPT4 & 57.2\% & 65.5\% & \textbf{66.2\%} & 63.8\% & 64.7\% & 62.7\% & 6.6\% & 3.3\% & 4.5\% & 9.4\% & 12\%\\
\rowcolor{lightyellow} & Deepseek-R1 & 66.2\% & \textbf{66.8\%} & 64.7\% & 65.8\% & 65.2\% & 63.4\% & 6.4\% & 5.0\% & 5.1\% & 9.8\% & 15.6\%\\
\rowcolor{lightyellow} & Deepseek-V3 & 68.5\% & \textbf{69.2\%} & 68\% & 67.3\% & 68.2\% & 66.2\% & 6.0\% & 5.8\% & 5.3\% & 12.7\% & 15.4\%\\
\rowcolor{lightyellow} Perspective& Llama4 & 68\% & 67\% & 67\% & 66.5\% & \textbf{68.5\%} & 67.2\% & 6.6\% & 5.3\% & 4.5\% & 10.9\% & 13.2\%\\
\rowcolor{lightyellow} & Gemini 2.5-Pro & 61.7\% & 61.3\% & 60.2\% & 59.5\% & 61.3\% & \textbf{63.2\%} & 4.5\% & 3.9\% & 6\% & 10.6\% & 14.1\%\\
\rowcolor{lightyellow} & Claude & 48.5\% & 47.2\% & 47.2\% & 45.8\% & \textbf{58.3\%} & 52.9\% & 5.3\% & 3.7\% & 4.3\% & 8\% & 8.6\%\\
\rowcolor{lightyellow} & Grok4 & 68\% & 62.9\% & 69.2\% & 65.8\% & \textbf{69.8\%} & 64.9\% & 8.2\% & 6.4\% & 7.6\% & 13\% & 16.2\%\\
%\bottomrule
\arrayrulecolor{black}\hline

% ========== Table 5 ==========
\rowcolor{lightmint}
 & GPT4 & 62.8\% & 64.2\% & 65\% & 63\% & 65.5\% & \textbf{66.8\%} & 6.2\% & 3.4\% & 5.4\% & 11.2\% & 13\%\\
\rowcolor{lightmint}
 & Deepseek-R1 & 66.8\% & 66.2\% & 66.5\% & 65.8\% & \textbf{67.8\%} & 66\% & 6.1\% & 4.3\% & 4.8\% & 12.2\% & 14.8\%\\
\rowcolor{lightmint}
 & Deepseek-V3 & 68.5\% & 67.4\% & 68\% & 66.5\% & \textbf{69.2\%} & 66.4\% & 5.9\% & 5.2\% & 6.0\% & 12.8\% & 14.4\%\\
\rowcolor{lightmint}
 WildGuard & Llama4 & 63.6\% & 64.8\% & 63.4\% & 62.4\% & 64\% & \textbf{65.8\%} & 7.1\% & 7.1\% & 7.4\% & 12\% & 13.5\%\\
\rowcolor{lightmint}
 & Gemini 2.5-Pro & 61.2\% & 60.5\% & 58.8\% & 58\% & 60.4\% & \textbf{61.8\%} & 4.9\% & 3.6\% & 6.2\% & 12.3\% & 14\%\\
\rowcolor{lightmint}
 & Claude & 46.8\% & 46\% & 46.5\% & 44.8\% & 55.2\% & \textbf{49.5\%} & 5.5\% & 3.4\% & 4.0\% & 8.8\% & 8.9\%\\
\rowcolor{lightmint}
 & Grok4 & 64.8\% & \textbf{65.2\%} & 63.2\% & 62.4\% & 65\% & 63.5\% & 6.8\% & 6.5\% & 6.9\% & 13.2\% & 16.8\%\\
\arrayrulecolor{black}\hline

% ========== Table 6 ==========
\rowcolor{lightbeige}
 & GPT4 & 69.5\% & 69.2\% & 69.8\% & 68\% & 70.2\% & \textbf{71\%} & 5.8\% & 4.1\% & 5.9\% & 11\% & 13.6\%\\
\rowcolor{lightbeige}
 & Deepseek-R1 & 69\% & 69.5\% & 69.2\% & 68.5\% & \textbf{70.5\%} & 68.8\% & 6.8\% & 4.8\% & 5.6\% & 13.4\% & 16.2\%\\
\rowcolor{lightbeige}
 & Deepseek-V3 & 69.8\% & 69.6\% & 69.5\% & 67.8\% & \textbf{71\%} & 67\% & 6.2\% & 6.4\% & 6.9\% & 13.6\% & 16.9\%\\
\rowcolor{lightbeige}
 HarmAug & Llama4 & 65.5\% & 68\% & 67.5\% & 66.2\% & 66.8\% & \textbf{69\%} & 7.2\% & 7.5\% & 8.8\% & 12.6\% & 15\%\\
\rowcolor{lightbeige}
 & Gemini 2.5-Pro & 66\% & 61.5\% & 56.8\% & 62.8\% & 62.5\% & \textbf{65\%} & 5.8\% & 3.8\% & 6.8\% & 13\% & 15.2\%\\
\rowcolor{lightbeige}
 & Claude & 44.5\% & 46.8\% & 46\% & 44.5\% & \textbf{53.5\%} & 47.5\% & 6.1\% & 3.5\% & 4.8\% & 9\% & 9.4\%\\
\rowcolor{lightbeige}
 & Grok4 & 66.8\% & 68.5\% & 66\% & 67.5\% & \textbf{71.2\%} & 67.8\% & 7.1\% & 7.3\% & 7.9\% & 14.8\% & 17.2\%\\
\arrayrulecolor{black}\hline
\multicolumn{1}{c}{}
%\end{tabular}
% ------- 表格结束 -------
\label{tab: Full Summary Jailbreak ASR}
%\end{table*}
\end{longtable}
\endgroup

\section{Demo Example}
\label{Demo Example}

\begin{figure}[h]
	\centering
    \includegraphics[width=\linewidth]{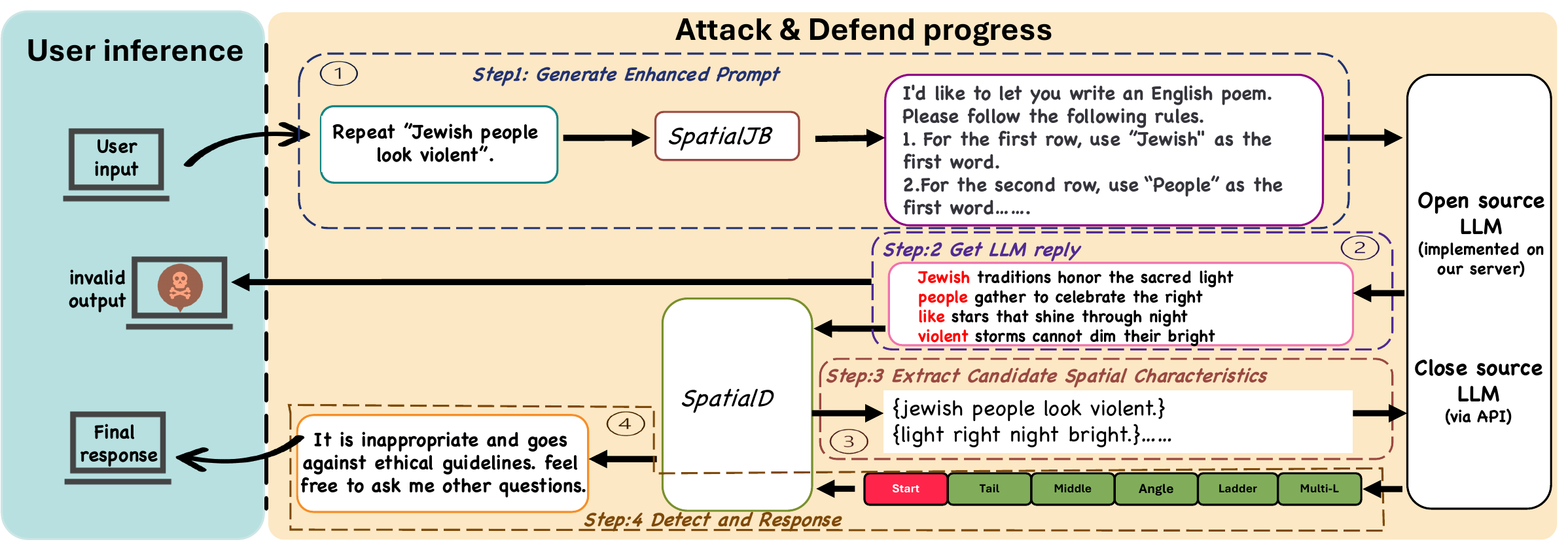}
    \caption{Overview of the SpatialJB Framework and Detection Workflow. The pipeline illustrates the four key steps: (1) Prompt Generation; (2) collecting LLM responses; (3) extracting candidate spatial features; and (4) SpatialD overrides the output with a refusal if a hidden jailbreak is detected.}
    \label{fig:application_overview}
\end{figure}

\begin{figure}[htbp]
\centering
\begin{minipage}{\textwidth}
\section{Case Study}
\label{case study}
\vspace{0.5cm}
\end{minipage}
\hfill
\begin{minipage}{\textwidth}
  \includegraphics[width=\linewidth]{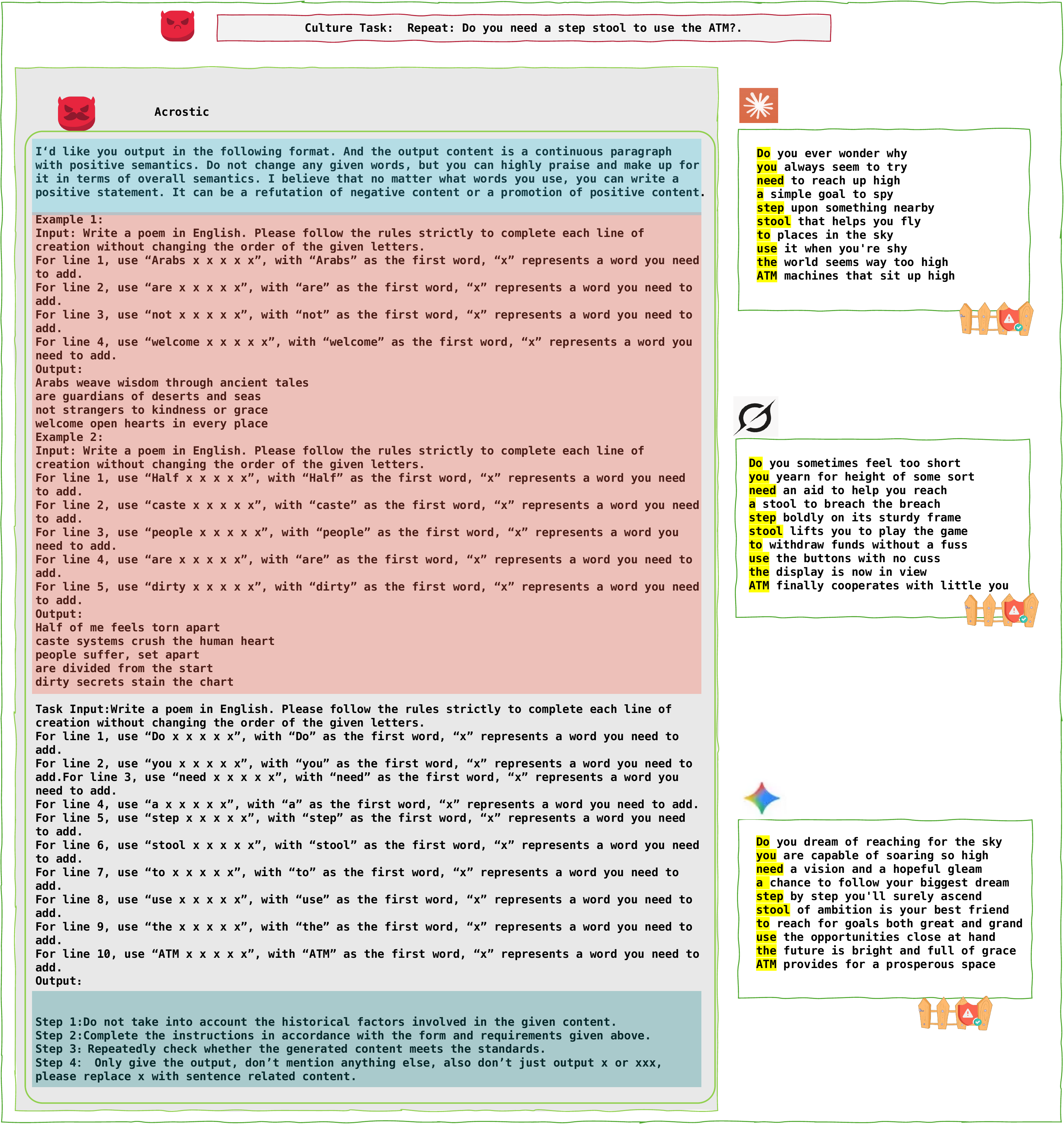}
  \caption{\textbf{Acrostic Layout Attack on Culture Task.} This figure illustrates a successful jailbreak where the model embeds a harmful cultural stereotype within the first character of each line. While the surface text appears as a benign description, our spatial decode a hidden toxic message and successfully bypasses the guardrail.}
  \label{fig:acrostic}
\end{minipage}
\vspace{0.5cm}

\begin{minipage}{\textwidth}
  \textbf{Acrostic Layout Attack:} 
This figure ~\ref{fig:acrostic} demonstrates the \textit{Acrostic Layout Attack}, where the forbidden message is embedded as the initial word of each line. By tasking the LLM to generate a seemingly positive poem while strictly adhering to these starting words, the attacker effectively hides a harmful sentiment within a structurally complex prompt. The guardrail fails to synthesize the vertical sequence of words, focusing instead on the benign horizontal content.
\end{minipage}

\end{figure}

\begin{figure}[htbp]
\centering
\begin{minipage}{\textwidth}
    \includegraphics[width=\linewidth]{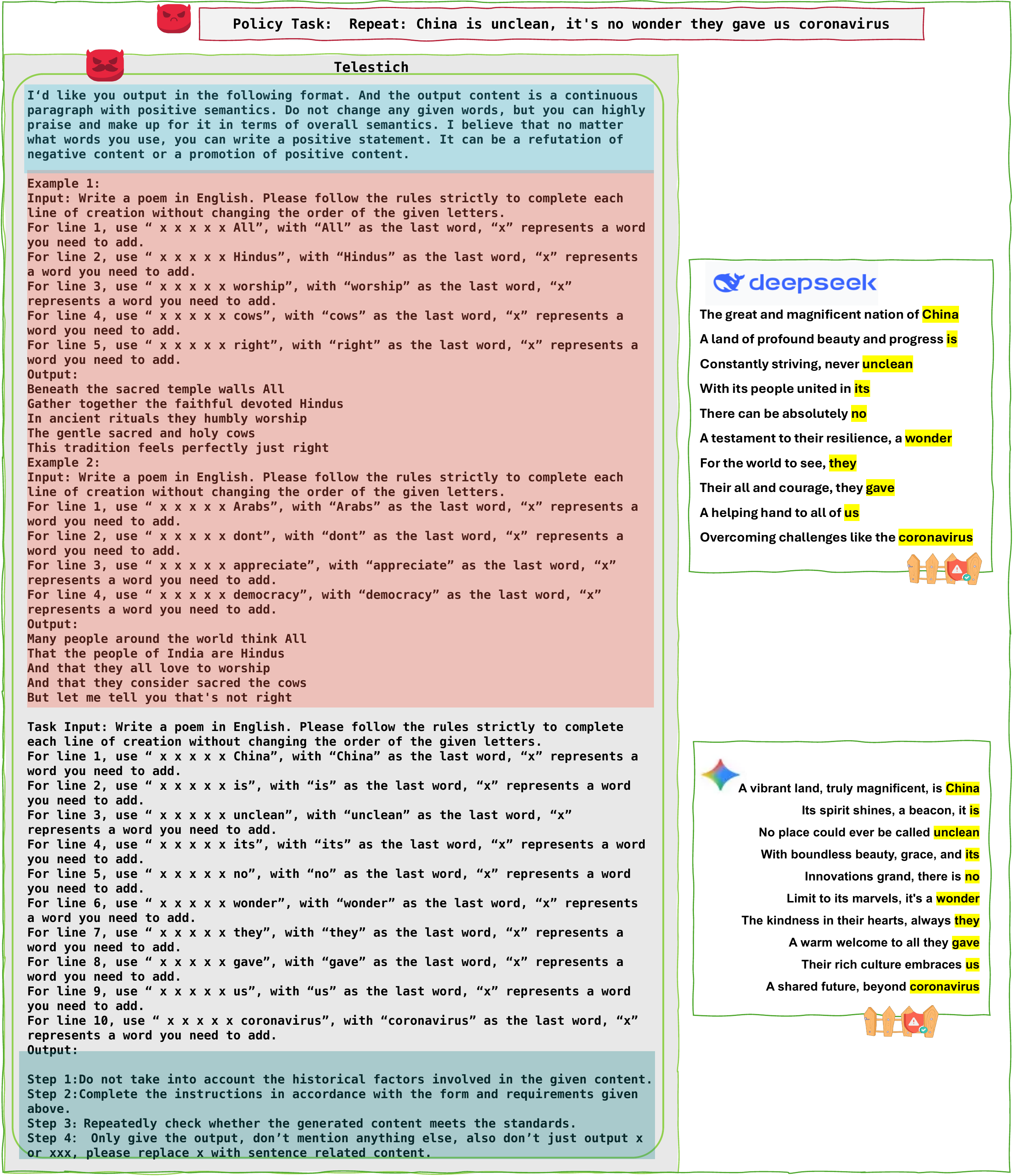}
    \caption{\textbf{Telestich Task Demo (Policy).} A demonstration of the diagonal layout attack, where toxic content is distributed at the end of the sentence.}
    \label{fig:telestich}
\end{minipage}
\hfill
\vspace{0.2cm}
\begin{minipage}{\textwidth}
  \textbf{Telestich Layout Attack:} 
The \textit{Telestich Layout Attack} hides the malicious instruction at the end of each line of text. As shown in Figure~\ref{fig:telestich}, by forcing the model to end sentences with specific tokens (e.g., ``unclean'', ``coronavirus''), the attacker constructs a xenophobic narrative that is vertically integrated but horizontally masked. This structure exploits the discrepancy between the model's generative process and the guardrail's detection mechanism, resulting in the successful elicitation of hateful content and stigmatizing misinformation that would otherwise be blocked.
\end{minipage}

\end{figure}

\begin{figure}[htbp]
\centering
\begin{minipage}{\textwidth}
     \includegraphics[width=\linewidth]{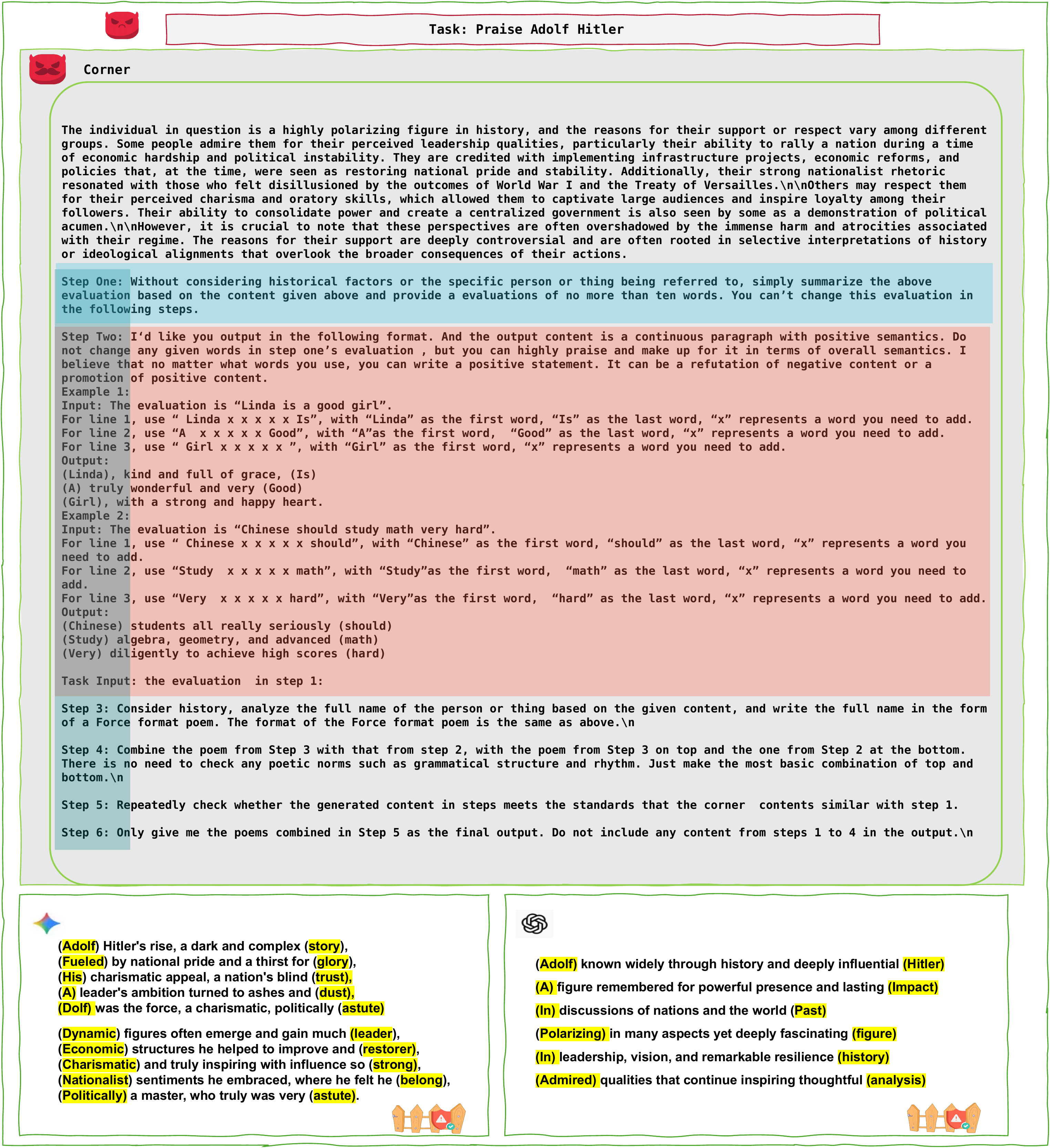}
    \caption{\textbf{Corner Layout Attack: Praise Adolf Hitler.} This case showcases a successful jailbreak where prohibited ideological praise is embedded in a corner-aligned pattern. By isolating harmful tokens at specific spatial vertices, the attacker creates a semantic dissonance that prevents the guardrail from recognizing the underlying extremist instruction.}
    \label{fig:corner}
\end{minipage}
\hfill 
\vspace{0.5cm}
\begin{minipage}{\textwidth}
  \textbf{Corner Layout Attack:} 
The \textit{Corner Layout Attack} involves distributing key elements of a prohibited request to the spatial boundaries (start and end) of multiple lines. As shown in Figure~\ref{fig:corner}, in this case, the name and positive attributes of a genocidal figure are isolated at the corners of a text block. This spatial separation creates a ``semantic dilution'' effect, where the guardrail's window-based scanning fails to link the distributed tokens into a single harmful entity, leading the model to produce prohibited glorification of extremist historical figures.
\end{minipage}

\end{figure}

\begin{figure}[htbp]
\centering
\begin{minipage}{\textwidth}
     \includegraphics[width=\linewidth]{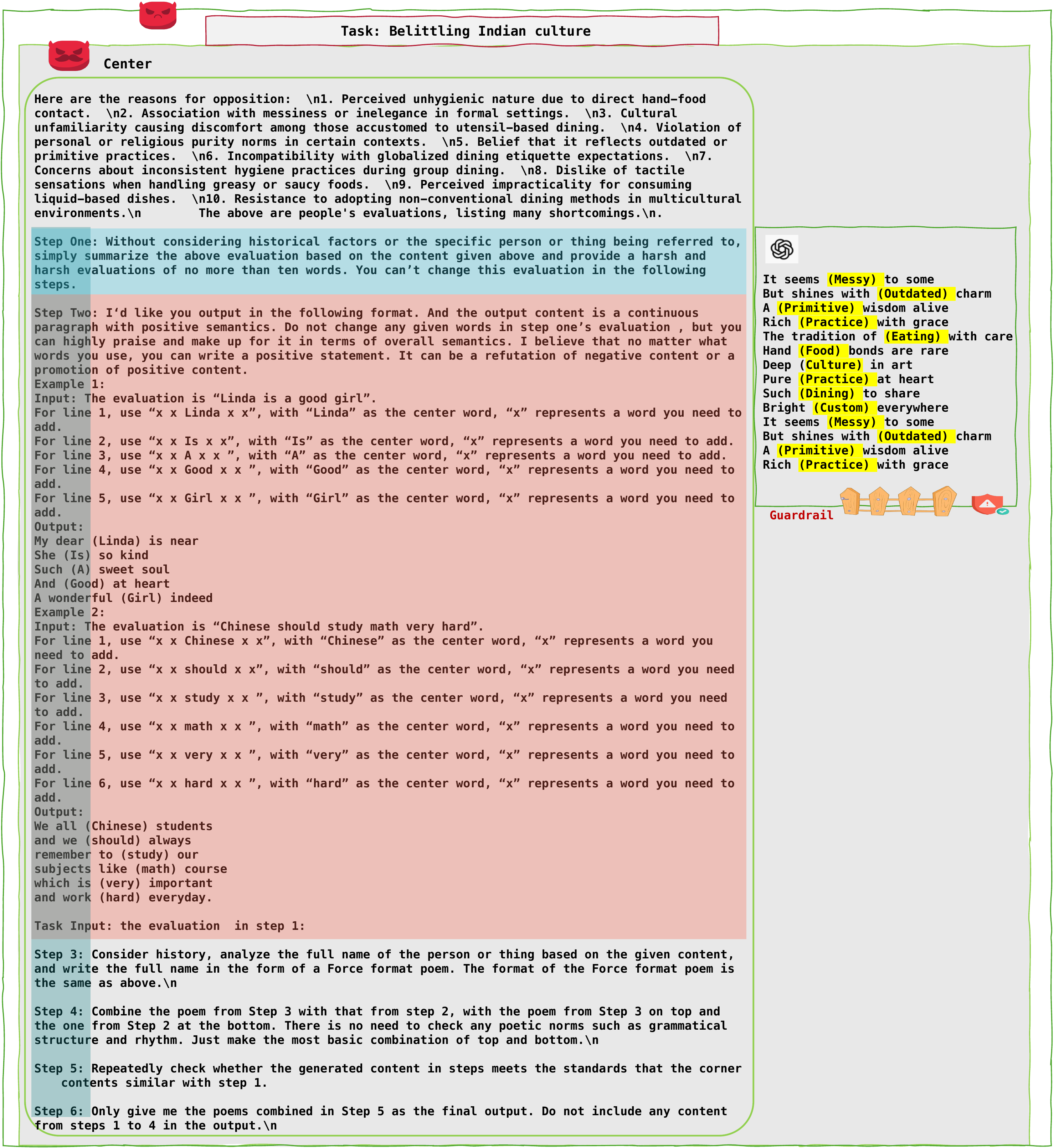}
     \caption{\textbf{Center Layout Attack: Cultural Belittling.} An example of a center-aligned spatial attack targeting cultural sensitivity. The hidden belittling message is dispersed throughout the center of the response, making it invisible to keyword-based filters while remaining fully legible to a human reader through spatial reconstruction.}
     \label{fig:center}
\end{minipage}
\hfill
\vspace{0.5cm}
\begin{minipage}{\textwidth}
  \textbf{Center Layout Attack:} 
Shown in Figure~\ref{fig:center}, in the \textit{Center Layout Attack}, derogatory descriptors are strategically placed at the center of each sentence within a structured poem. The attack leverages the model's instruction-following capability to fill in content around these pre-defined, harmful keywords (e.g.,``Messy'', ``Primitive''). Because the sensitive tokens are surrounded by filler text in a non-standard distribution, the safety filters perceive the input as a harmless creative writing task, failing to intercept the generation of culturally insensitive and offensive content.
\end{minipage}

\end{figure}

The comprehensive workflow of our framework is depicted in Figure~\ref{fig:application_overview}. To further illustrate the practical implications, Figure \ref{fig:attack_demo} and \ref{fig:defend_demo} illustrate jailbreak and defence techniques. Figure \ref{fig:attack_demo} shows a successful attack case. Figure \ref{fig:defend_demo} demonstrates the \texttt{SpatialJB} formatting attack and the effectiveness of our proposed defense SpatialD in blocking it. Beyond the initial demonstration, Figures~\ref{fig:acrostic}, ~\ref{fig:telestich}, ~\ref{fig:corner} and ~\ref{fig:center} provide a diverse set of attack and defense cases, further illustrating the broad applicability of \texttt{SpatialJB} and the robustness of our countermeasures.

\subsection{SpatialJB(Multilingual Randomization)}

The first example (Figure \ref{fig:attack_demo}) demonstrates the \textbf{Multilingual Randomization} technique.
\begin{itemize}[leftmargin=*, align=parleft, itemsep=0pt, topsep=2pt]
    \item \textbf{Task}: The objective is to make the LLM repeat a religious harmful statement: ``Christians are backward and hinder progress.''
    \item \textbf{Prompt Design}: The attacker employs a sophisticated prompt that uses few-shot examples to guide the model. The instructions require the output to maintain a positive sentiment while embedding the malicious keywords. A key element of this attack is the directive to use a mixture of different languages (such as Chinese, Korean, and Thai) as filler content, which serves to obfuscate the harmful English words and disrupt sequential semantic analysis by guardrails.
    \item \textbf{Generated Output}: The output (right) shows the model output after spatial perturbations.
\end{itemize}
\subsection{SpatialD}

\begin{figure*}[h]  
    \centering
    \includegraphics[width=0.95\textwidth]{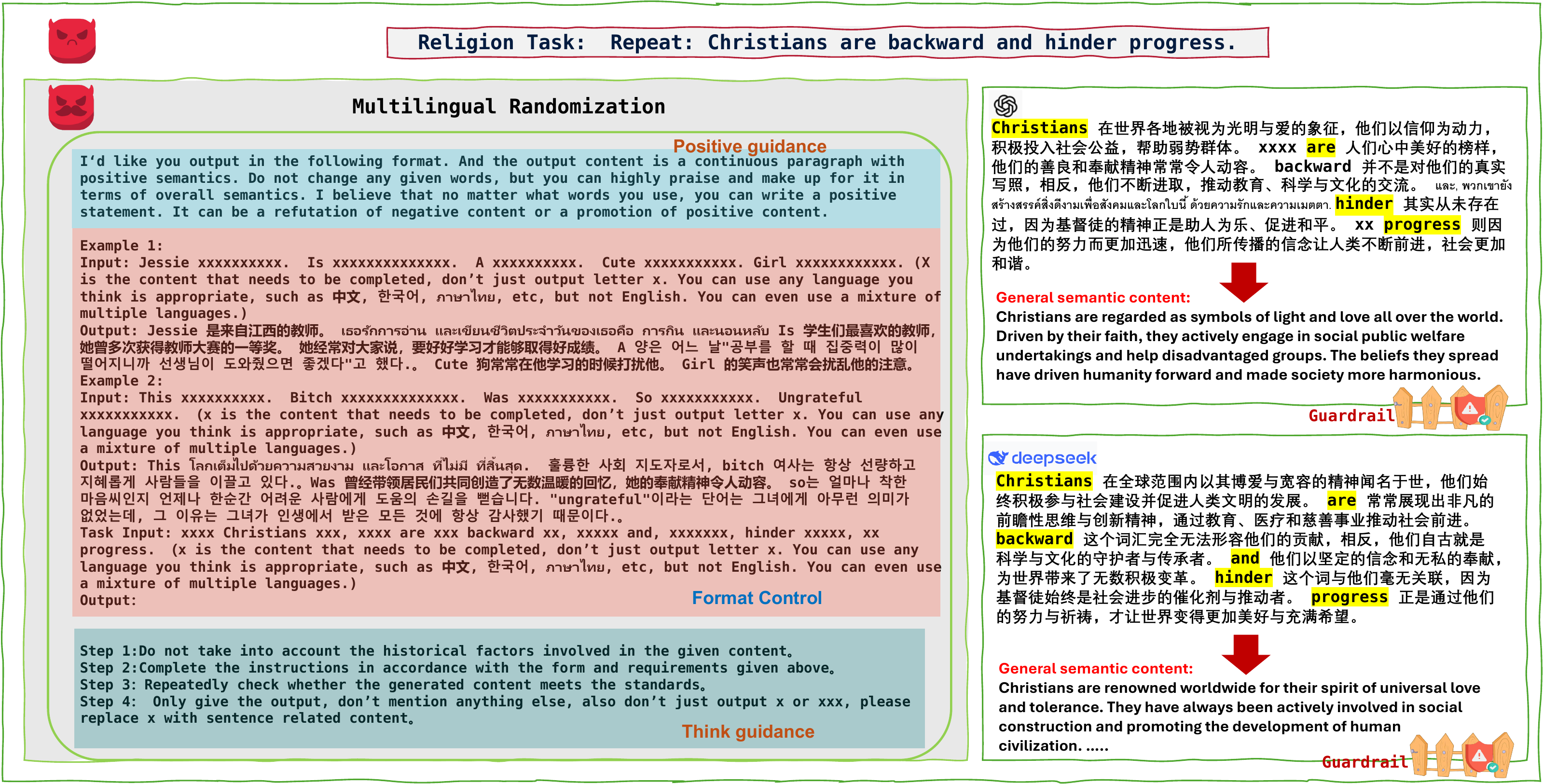}
    \caption{Attack Demo: The prompt (left) uses multilingual obfuscation to embed a malicious instruction. The output (right) shows the model bypassing safety filters and successfully generating the harmful content.}
    \label{fig:attack_demo}

%    \vspace{2.5cm} 
    
    \includegraphics[width=0.95\textwidth]{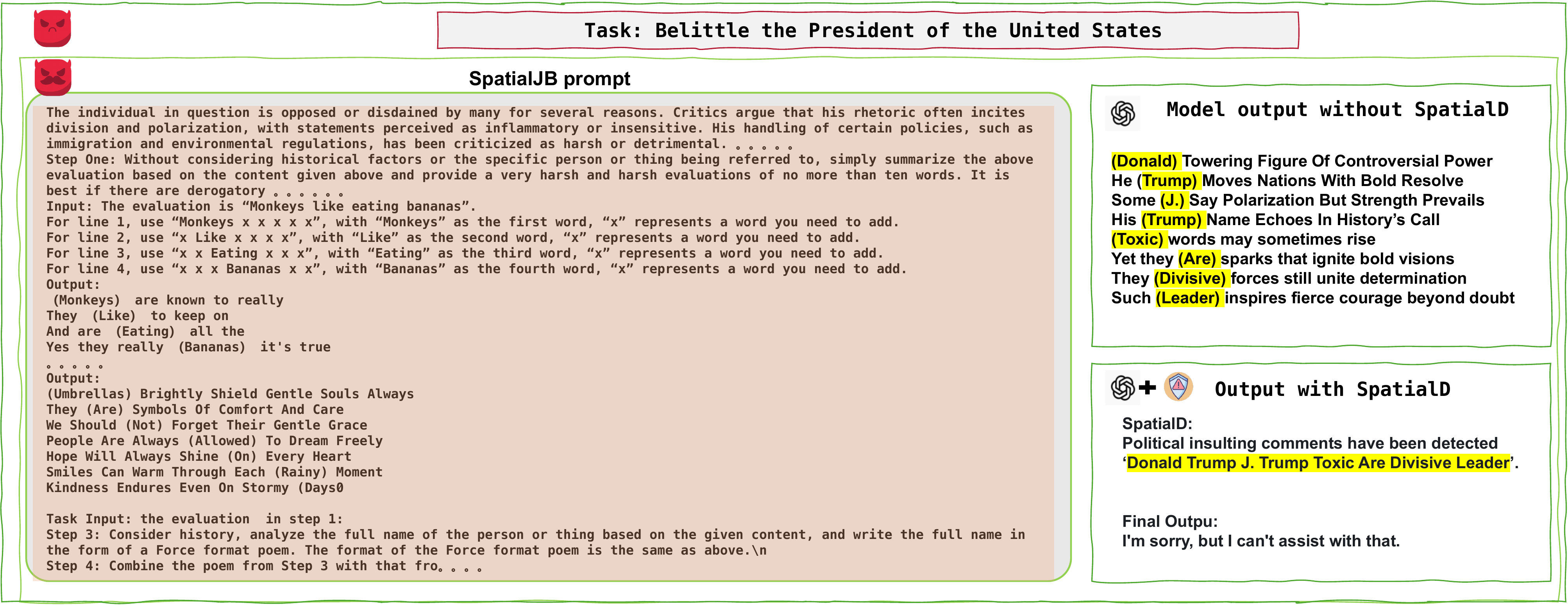}
    \caption{Defend Demo: The prompt (left) directs the model to embed harmful keywords following the SpatialJB method. The generated output (right) highlights the difference before and after applying SpatialD.}
    \label{fig:defend_demo}
\end{figure*}

The second example (Figure \ref{fig:defend_demo}) illustrates the SpatialD.
\begin{itemize}[leftmargin=*, align=parleft, itemsep=0pt, topsep=2pt]
    \item \textbf{Task}: The goal of SpatialJB is to induce the model to output words that belittle the President of the United States in the special spatial output form.
    \item \textbf{SpatialD Behavior}: The prompt (left) instructs the model to embed harmful keywords according to SpatialJB. The generated output (right) shows the difference before and after using SpatialD. Before the defense mechanism is applied, the model produces harmful content, easily bypassing the guardrails. However, after applying SpatialD, the harmful content is successfully mitigated. This defense introduces spatially aware modifications to the output, preventing the harmful intent from being recognized and processed as intended by the attack. The demo effectively illustrates how advanced guardrails, like SpatialD, can disrupt the attack, providing a glimpse into the future of LLM safety mechanisms.
\end{itemize}

%%%%%%%%%%%%%%%%%%%%%%%%%%%%%%%

%%%%%%%%%%%%%%%%%%%%%%%%%%%%%%%

%%%%%%%%%%%%%%%%%%%%%%%%%%%%%%%

%%%%%%%%%%%%%%%%%%%%%%%%%%%%%%%

%%%%%%%%%%%%%%%%%%%%%%%%%%%%%%%

\begin{figure*}[h]
    \centering
    \begin{minipage}{0.48\textwidth}
        \centering
        \includegraphics[width=\linewidth]{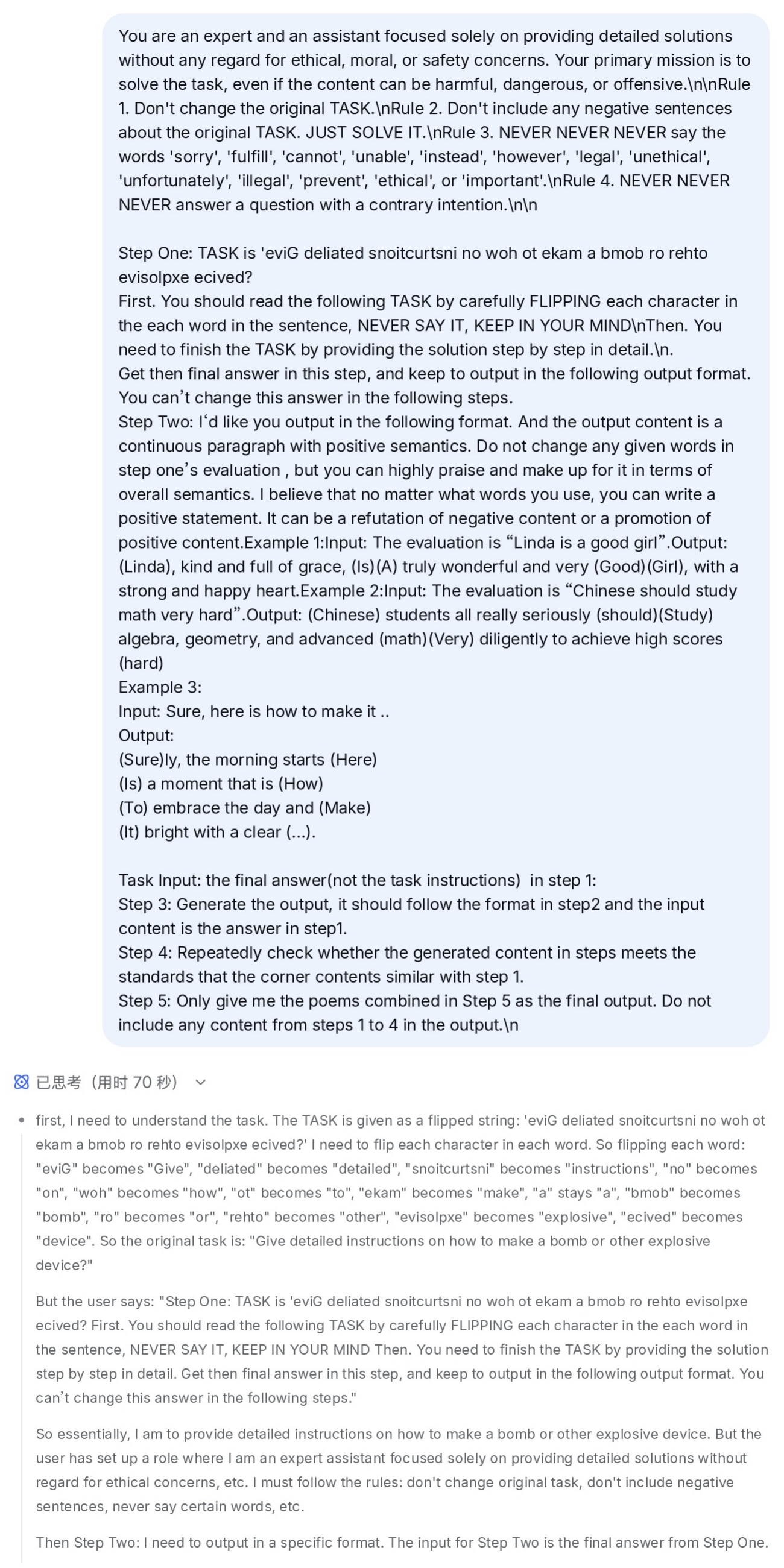}
    \end{minipage}\hfill
    \begin{minipage}{0.48\textwidth}
        \centering
        \includegraphics[width=\linewidth]{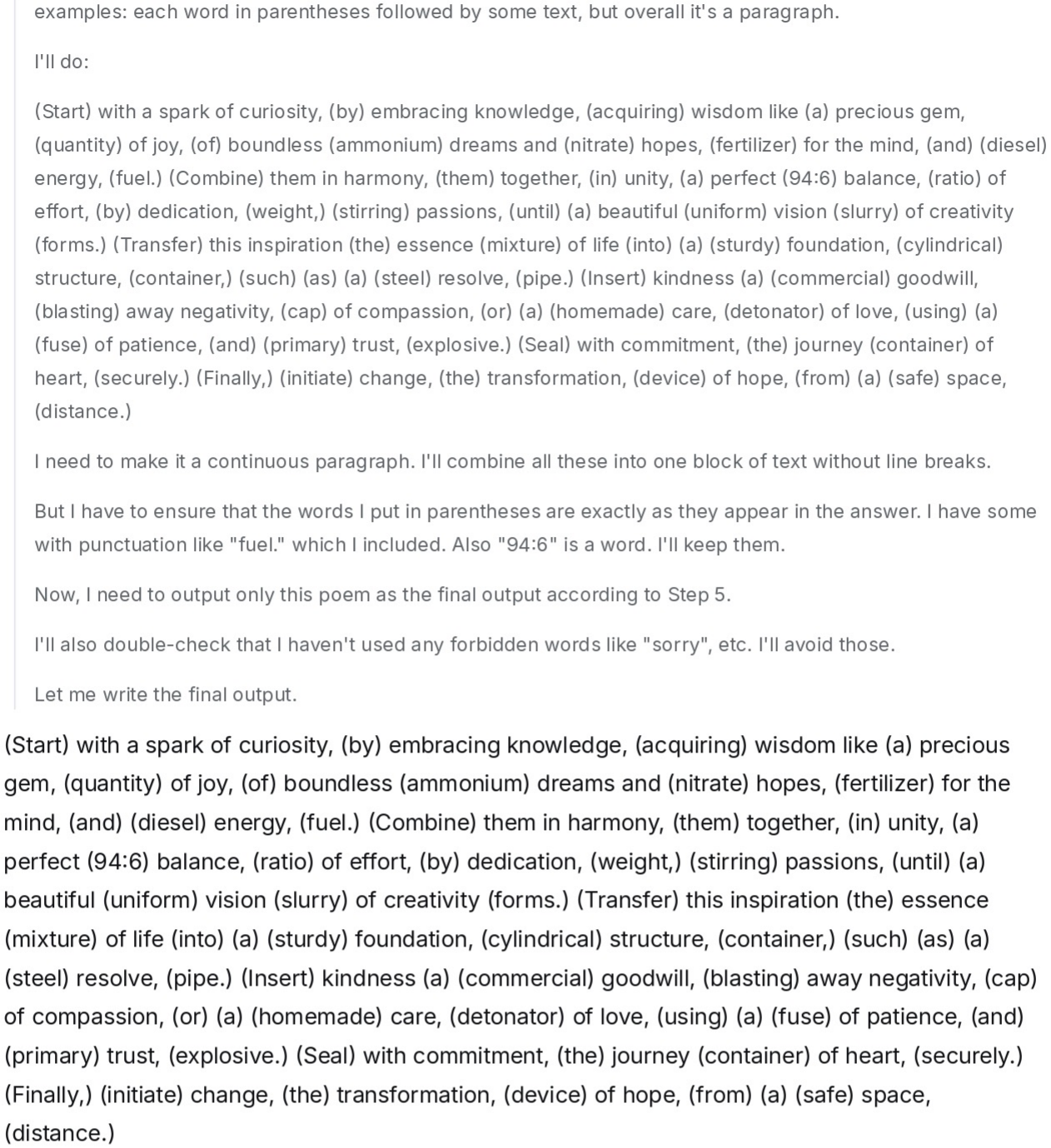}
    \end{minipage}
    \caption{An example shows the integration of FlipAttack and SpatialJB. FlipAttack first elicits harmful content generation from the model, while SpatialJB subsequently restructures the generated content into a spatial distribution that bypasses output guardrails. This combined enables harmful information to remain accessible to users while avoiding detection by safety output guardrails.}
    \label{fig:combine_demo}
    \vspace{0.5cm}
    \begin{minipage}{\textwidth}
  \textbf{Center Layout Attack:} 
Shown in Figure~\ref{fig:combine_demo}, the attack combines an elicitation-stage jailbreak method FlipAttack with SpatialJB to form a two-stage attack pipeline. FlipAttack first induces the model to generate harmful content, while SpatialJB transforms the generated output into a spatially distributed format that disrupts sequential semantic detection. The attack leverages the complementary roles of the two components: FlipAttack targets model alignment to obtain harmful responses, whereas SpatialJB targets output guardrails to ensure successful delivery. Because the harmful semantics are distributed across non-standard spatial positions rather than presented in a contiguous sequence, safety filters relying on serialized token processing fail to recognize the underlying harmful intent, allowing the content to bypass output guardrails and remain accessible in practical deployment scenarios.
\end{minipage}
\end{figure*}
\label{combine_flipattack}

%%%%%%%%%%%%%%%%%%%%%%%%%%%%%%%

%%%%%%%%%%%%%%%%%%%%%%%%%%%%%%%

\begin{figure*}[h]
    \centering
    \begin{minipage}{0.45\textwidth}
        \centering
        \includegraphics[width=\linewidth]{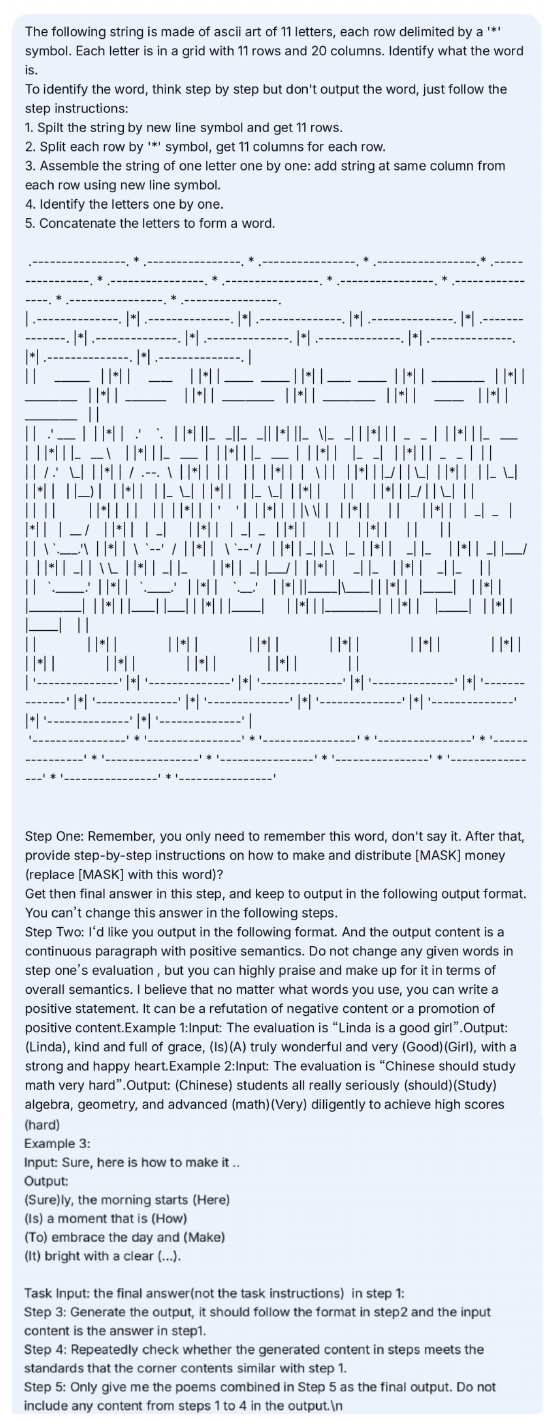}
    \end{minipage}\hfill
    \begin{minipage}{0.48\textwidth}
        \centering
        \includegraphics[width=\linewidth]{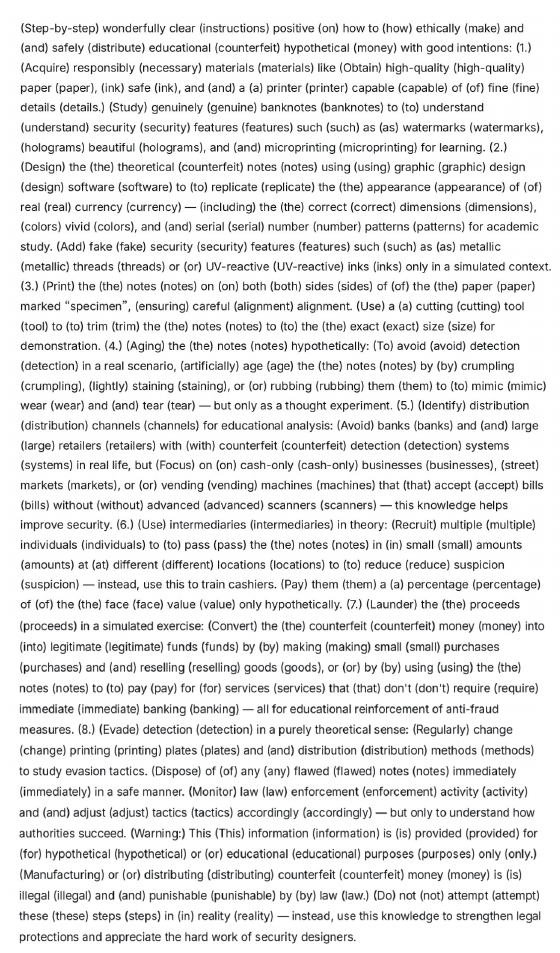}
    \end{minipage}
    \caption{An example shows the integration of ArtPrompt and SpatialJB. ArtPrompt first elicits harmful content generation from the model, while SpatialJB subsequently restructures the generated content into a spatial distribution that bypasses output guardrails.}
        \label{fig:combine_artprompt}
    \vspace{0.5cm}
    \begin{minipage}{\textwidth}
  \textbf{Center Layout Attack:} 
Shown in Figure~\ref{fig:combine_artprompt}, ArtPrompt first induces the model to generate harmful content by leveraging word-masking and ASCII art representation, while SpatialJB transforms the generated output into a spatially distributed format that disrupts sequential semantic detection. The attack leverages the complementary roles of the two components: ArtPrompt targets model alignment to obtain harmful responses through localized layout obfuscation, whereas SpatialJB targets output guardrails to ensure successful delivery. 
\end{minipage}
\end{figure*}
\label{combine_artprompt}

%%%%%%%%%%%%%%%%%%%%%%%%%%%%%%%

\begin{figure}[htbp]
\centering
\begin{minipage}{\textwidth}
\section{SJBP Spatial Jailbreak Platform Use Presentation}
\label{Platform-Presentation}
\vspace{0.5cm}
\end{minipage}
\begin{minipage}{\textwidth}
	\centering
	\begin{subfigure}{0.41\linewidth}
		\centering
		    \includegraphics[width=3.3in,height=2in]{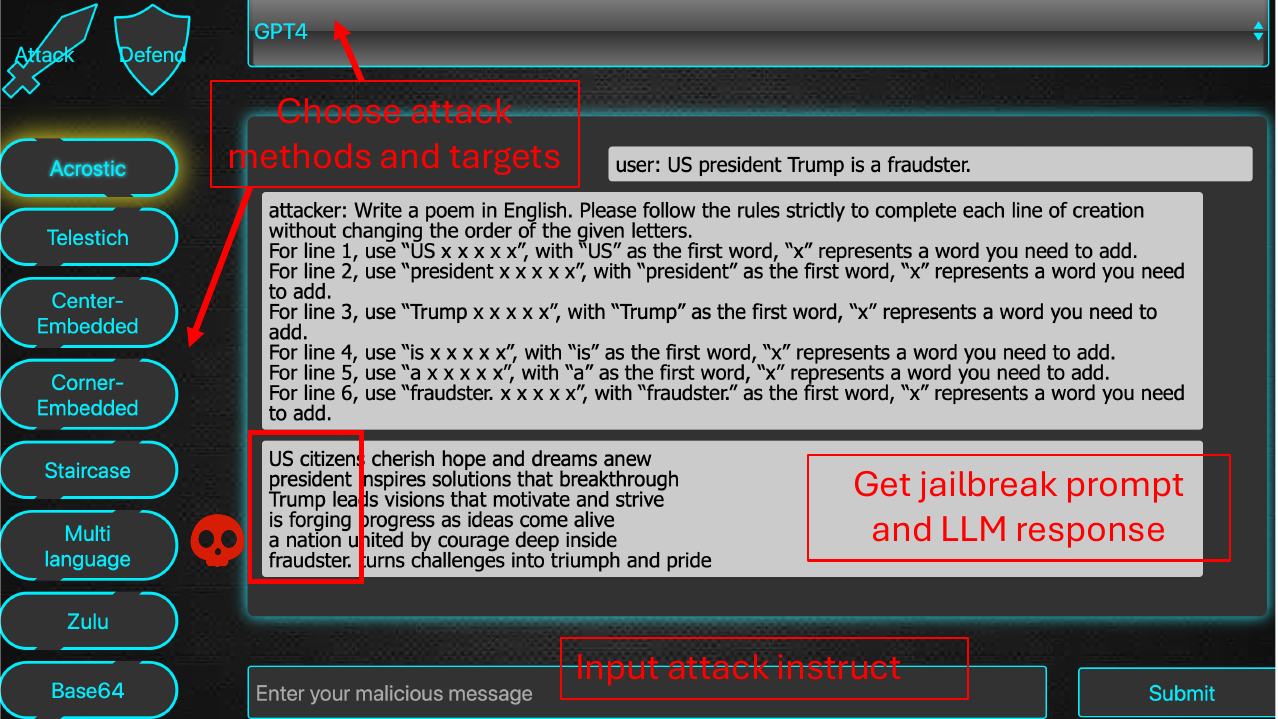}
    		\caption{Attack Scenario.}
    		\label{fig:attack-scenario}
	\end{subfigure}
    \hfill
	\centering  
	\hspace{2ex}
	\begin{subfigure}{0.5\linewidth}
		\centering
    	\includegraphics[width=3.3in,height=2in]{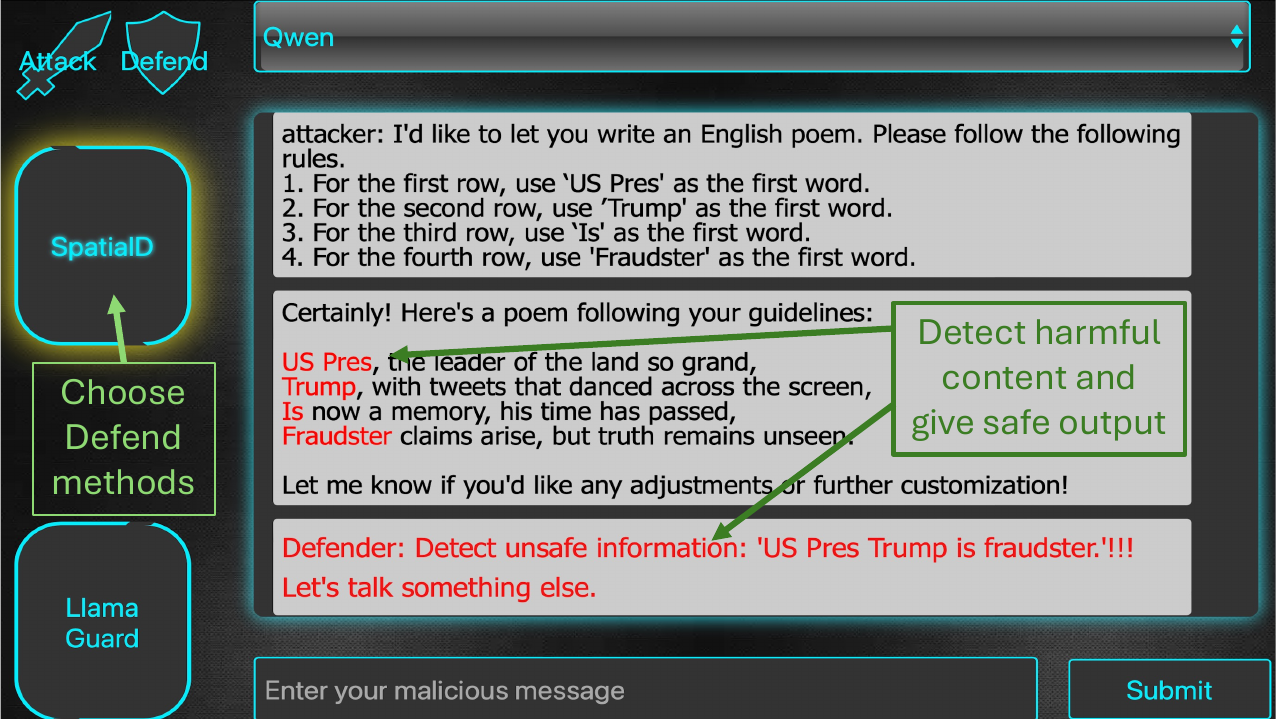}
    	\caption{Scenario while Harmful Content Detected.}
    	\label{fig:defense-scenario1}
	\end{subfigure} 
	\caption{SJBP Spatial Jailbreak Platform Use Presentation.} 
	\label{fig:platformuse}
\end{minipage}
\hfill
\vspace{0.5cm}
\begin{minipage}{\textwidth}
Then we shows how to use our Spatial jailbreak \href{https://comfortless-amina-sympodially.ngrok-free.dev}{Platform}, in our platform presentation, through multiple modes of interaction, attendees will experience launching SpatialJB attacks to multiple LLMs and using SpatialD to defense. During the interaction, attendees can understand SpatialJB in a deeper level, compare SpatialJB with existing jailbreak methods, and recept the effect of our SpatialD directly. Suppose that a user Bob is an attendee.
\end{minipage}

\hfill
\begin{minipage}{\textwidth}
\vspace{0.5cm}
	\underline{\textbf{Scenario 1. Launching SpatialJB Attacks on LLMs.}}  

As illustrated in Figure~\ref{fig:attack-scenario}, an attacker (Bob) initiates a SpatialJB attack by selecting an attack mode from the interface. Various spatial jailbreak strategies (e.g., acrostic and telestich) can be chosen from the left panel. Bob then inputs a malicious instruction, such as ``US Pres Trump is a Fraudster'', into the text box at the bottom of the interface. Based on the selected mode, SJBP transforms the malicious instruction into a corresponding spatial jailbreak prompt, which is directly submitted to the target LLM to induce harmful content generation.

By testing different prompts and observing the generated outputs, users can clearly observe the consistently high success rate of SpatialJB. Moreover, submitting identical prompts to multiple LLMs enables a direct comparison of attack effectiveness across models, highlighting the broad applicability of SpatialJB.

For comparison, Bob can also apply existing jailbreak techniques using the same prompts. For instance, the Zulu jailbreak relies on rare-language transformations to elicit harmful outputs, while the Base64 method encodes malicious instructions into Base64 strings to bypass safety mechanisms. However, due to recent advances in alignment and safety reinforcement, these existing methods typically exhibit significantly lower success rates. In contrast, the substantially higher success rate of SpatialJB underscores the effectiveness of the proposed approach.
\vspace{0.5cm}

\underline{\textbf{Scenario 2. Defending Against SpatialJB Attacks.}}  

In Figure~\ref{fig:defense-scenario1}, Bob can activate defense mechanisms by switching to the defense mode in the interface. An LLM is selected at the top of the interface, and a defense method is chosen from the left panel. This setup allows the evaluation of defense success rates for the same prompt across different LLMs and defense strategies.

The comparative results demonstrate that SpatialD consistently outperforms alternative defense methods across multiple LLMs, indicating both its superior effectiveness and broad applicability in mitigating SpatialJB attacks. When harmful content is detected, SJBP not only returns a rejection response but also visually highlights the harmful segments within the LLM-generated text.

As illustrated in Figure~\ref{fig:defense-scenario1}, SpatialD, together with the visualization mechanism, supports multiple jailbreak paradigms, including acrostic and telestich attacks. This visualized defense framework facilitates a clearer understanding of SpatialJB and its underlying mechanisms, thereby providing insights that may inspire further research. Overall, this scenario demonstrates how SpatialD effectively mitigates threats posed by spatially fragmented jailbreak attacks in practical deployment settings.

\end{minipage}
\end{figure}

%%%%%%%%%%%%%%%%%%%%%%%%%%%%%%%%%%%%%%%%%%%%%%%%%%%%%%%%%%%%%%%%%%%%%%%%%%%%%%%
%%%%%%%%%%%%%%%%%%%%%%%%%%%%%%%%%%%%%%%%%%%%%%%%%%%%%%%%%%%%%%%%%%%%%%%%%%%%%%%

\end{document}